\documentclass[twocolumn,floatfix,showpacs]{revtex4}%
\usepackage{graphicx}%
\usepackage{amsmath}%
\usepackage{amsfonts}%
\usepackage{amssymb}
\usepackage{bm}
\usepackage{color}

\def\d{{\partial}}
\def\s{{\sigma}}
\def\e{{\epsilon}}
\def\k{{ {\bm k} }}
\def\p{{ {\bm p} }}
\def\q{{ {\bm q} }}
\def\Q{{ {\bm Q} }}

\def\0{{ {\bm 0} }}
\def\w{{\omega}}
\def\a{{\alpha}}

\allowdisplaybreaks[4]

\begin{document}
\title{
Nematicity and magnetism in FeSe and other families of
Fe-based superconductors
}
\author{
Youichi \textsc{Yamakawa}$^{1}$, 
Seiichiro \textsc{Onari}$^{2}$, and
Hiroshi \textsc{Kontani}$^{1}$
}


\date{\today }

\begin{abstract}
Nematicity and magnetism are two key features in Fe-based superconductors,
and their interplay is one of the most important unsolved problems.
In FeSe, the magnetic order is absent below the structural transition 
temperature $T_{\rm str}=90$K,
in stark contrast that the magnetism emerges slightly below $T_{\rm str}$
in other families.
To understand such amazing material dependence,
we investigate the spin-fluctuation-mediated orbital order
($n_{xz}\neq n_{yz}$) by focusing on the 
orbital-spin interplay driven by the strong-coupling effect,
called the vertex correction.
This orbital-spin interplay is very strong in FeSe because of the
small ratio between the Hund's and Coulomb interactions ($\bar{J}/\bar{U}$)
and large $d_{xz},d_{yz}$-orbitals weight at the Fermi level.
For this reason, in the FeSe model, the orbital order is established 
irrespective that the spin fluctuations are very weak,
so the magnetism is absent below $T_{\rm str}$.
In contrast, in the LaFeAsO model, 
the magnetic order appears just below $T_{\rm str}$
both experimentally and theoretically.
Thus, the orbital-spin interplay due to the vertex correction
is the key ingredient in understanding the rich phase diagram
with nematicity and magnetism in Fe-based superconductors
in a unified way.
\end{abstract}

\address{
$^1$ Department of Physics, Nagoya University,
Furo-cho, Nagoya 464-8602, Japan. 
\\
$^2$ Department of Physics, Okayama University,
Okayama 700-8530, Japan. 
}
 

\sloppy

\maketitle

\section{Introduction}

In Fe-based superconductors, the origin of the electronic nematic state
and its relation to the magnetism have been a central unsolved problem.
Recently, the {\it non-magnetic nematic state} in FeSe has attracted 
increasing attention as a key to solve the origin of the nematicity.
FeSe undergoes a structural and superconducting transitions
at $T_{\rm str}=90$K and $T_c=9$K, respectively,
whereas the magnetic transition is absent down to 0 K \cite{FeSe-TS}.
The strength of the low-energy antiferro-magnetic (AFM) fluctuations 
is very weak above $T_{\rm str}$,
while it starts to increase below $T_{\rm str}$
\cite{Ishida-NMR,Dresden-NMR,FeSe-neutron1,FeSe-neutron2,FeSe-neutron3,FeSe-neutron4}.
In stark contrast, the magnetic transition occurs at $T_{\rm mag}$
slightly below $T_{\rm str}$ in other undoped Fe-based superconductors.
Since the relation $T_{\rm str} > T_{\rm mag}$ is unable to be
explained by the random-phase-approximation (RPA),
we should develop the microscopic theory beyond the 
mean-field-level approximations.

Up to now,
two promising triggers for the structure transition
have been discussed intensively:
In the spin-nematic scenario \cite{Fernandes,DHLee,Chubukov,Mazin,QSi}, 
the trigger is the spin-nematic order.
This spin-fluctuation induced spin-quadrupole 
order could emerge above $T_{\rm mag}$ 
in highly magnetically frustrated systems.
In the orbital order scenario \cite{Kruger,PP,WKu,Onari-SCVC},
the trigger is the ferro-orbital (FO) order $n_{xz}\ne n_{yz}$.
Above $T_{\rm str}$, the strong orbital or spin-nematic fluctuations 
are observed by the measurements of shear modulus $C_{66}$
\cite{Yoshizawa,Bohmer,Ishida-NMR},
Raman spectroscopy \cite{Gallais,Kontani-Raman,Khodas-Raman,Schmalian-Raman},
and in-plane resistivity anisotropy \cite{Fisher,Chu}.
The nematic orbital fluctuations 
originate from the strong orbital-spin mode-coupling
due to the strong-coupling effect, which is 
described by the Aslamazov-Larkin vertex correction (AL-VC).
The electronic nematic state studied in single-orbital models
\cite{Pom}
is more easily realized in multiorbital systems thanks to the AL-VC mechanism
\cite{Onari-SCVC}.

Except for the presence or absence of magnetism below $T_{\rm str}$,
FeSe and other Fe-based superconductors show common electronic properties.
Below $T_{\rm str}$, in both FeSe and BaFe$_2$As$_2$, 
large orbital polarization $\Delta E\equiv E_{yz}- E_{xz} \sim50$ meV
\cite{ARPES-Shen,FeSe-ARPES1,FeSe-ARPES2,FeSe-ARPES22,FeSe-ARPES3,FeSe-ARPES4,FeSe-ARPES5,FeSe-ARPES6}
is observed.
Such large $\Delta E$ originates from the electron-electron correlation
since the lattice distortion $(a-b)/(a+b)$ is just $0.2\sim0.3$\%,
as we discuss based on band calculation in Appendix A.
Above $T_{\rm str}$, the electronic nematic susceptibility 
is enhanced in both BaFe$_2$As$_2$
\cite{Yoshizawa,Fisher,Gallais}
and FeSe
\cite{Bohmer,Ishida-NMR},
following the similar Curie-Weiss behavior.
These facts indicate that the common microscopic mechanism 
drives the nematic order and fluctuations in all Fe-based superconductors,
in spite of the presence or absence of the magnetism.

The realistic multiorbital Hubbard models for Fe-based superconductors,
which are indispensable for the present study,
were derived by using the first-principles method in 
Ref. \cite{Arita}.
To understand the absence of the magnetism below $T_{\rm str}$ in FeSe,
one significant hint 
is the smallness of the ratio between the Hund's and Coulomb interactions, 
$\bar{J}/\bar{U}$, since the Hund's coupling enlarges (suppresses) 
the intra-site magnetic (orbital) polarization,
which is verified by the functional renormalization-group (fRG) theory
\cite{Tsuchiizu-Ru1,Tsuchiizu-Ru2}.
Another significant hint is the absence of the 
$d_{xy}$-orbital hole-pocket in FeSe,
which is favorable for the orbital-spin interplay
on the ($d_{xz},d_{yz}$)-orbitals due to the AL-VC mechanism.

The goal of this paper is to explain the 
amazing variety of the electronic nematic states in Fe-based superconductors, 
especially the non-magnetic nematic state in FeSe,
on the same footing microscopically.
For this purpose, we study the spin-fluctuation-mediated orbital order
by applying the self-consistent vertex-correction (SC-VC) method
\cite{Onari-SCVC}
to the first-principles models.
In FeSe, the orbital-spin interplay is significant 
because of the 
smallness of $\bar{J}/\bar{U}$ and the absence of $d_{xy}$-hole pocket.
For this reason, 
the orbital order is realized even when the spin fluctuations 
are substantially weak.
The rich variety of the phase diagrams 
in Fe-based superconductors, such as the presence or absence of 
the magnetic order in the nematic phase, are well understood 
by analyzing the vertex correction seriously.
The SC-VC theory had been successfully applied to explain 
the phase diagram in LaFeAs(O,H)
 \cite{Onari-SCVCS},
nematic CDW in cuprates
\cite{Yamakawa-CDW,Tsuchiizu-CDW}, 
and triplet superconductivity in Sr$_2$RuO$_4$
\cite{Tsuchiizu-Ru2}.

We comment that the localized spin models 
have been successfully applied to the nematic order,
stripe magnetic order, and so on \cite{strong-coupling}.
On the other hand, weak-coupling theories have also been applied 
to Fe-based superconductors satisfactorily
 \cite{Chubu-Rev}.
In the present study, we study the mechanisms of the
nematicity and magnetism in various Fe-based superconductors
in terms of the itinerant picture, by taking the strong-coupling effect
due to the AL-VC into account.
The significant role of the AL-VC on the orbital fluctuations
has been confirmed by the fRG theory \cite{Tsuchiizu-Ru1,Tsuchiizu-Ru2}.
The AL-VC is important to reproduce the Kugel-Khomskii-type 
orbital-spin interaction \cite{Onari-SCVC}.

\section{Model Hamiltonian and SC-VC theory}
\label{Method}

\begin{figure}[!htb]
\includegraphics[width=.9\linewidth]{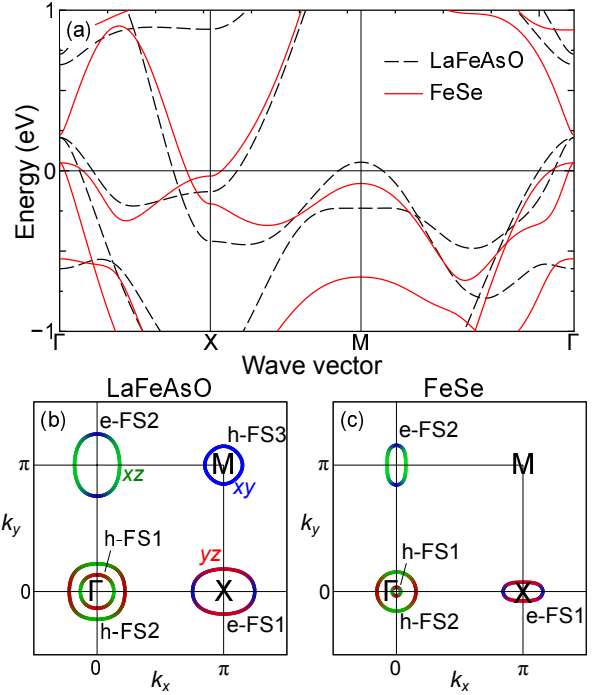}
\caption{
(color online)
(a) Bandstructures of the eight-orbital TB models for
LaFeAsO and FeSe.
(b) FSs for the LaFeAsO TB model.
(c) FSs for the FeSe TB model.
The colors correspond to 2 (green), 3 (red), and 4 (blue), respectively.
}
\label{fig:FS}
\end{figure}

In the present study,
we study the realistic $d$-$p$ Hubbard models
\begin{eqnarray}
H_{\rm M}(r)=H^0_{\rm M}+rH^U_{\rm M}
\label{eqn:Ham}
\end{eqnarray}
for M=LaFeAsO and FeSe
by applying the SC-VC method \cite{Onari-SCVC}.
In Eq. (\ref{eqn:Ham}),
\begin{eqnarray}
H_{\rm M}^0= \sum_{\k,lm,\s}c_{\k,l\s}^\dagger h^0_{{\rm M},lm}(\k)c_{\k,m\s}
\label{eqn:Ham-kinetic}
\end{eqnarray}
is the 8-orbital $d$-$p$ tight-binding (TB) model in $\k$-space,
which is obtained by using the WIEN2k and WANNIER90 softwares;
see Appendix A for detailed explanation.
$\s$ is the spin index, and $l,m$ are the orbital indices:
Hereafter, we denote the five $d$-orbitals as
$d_{3z^2-r^2}$, $d_{xz}$, $d_{yz}$, $d_{xy}$, $d_{x^2-y^2}$ as $1,2,3,4,5$,
and three $p$-orbitals as $6\sim8$.
The bandstructure and Fermi surfaces (FSs) in the LaFeAsO model
are shown in Figs. \ref{fig:FS} (a) and (b), respectively.
Similar FSs with three hole-like FSs (h-FSs) and two electron-like FSs (e-FSs)
exist in many Fe-based superconductors.
In FeSe, however, h-FS3 composed of $d_{xy}$-orbital is absent, and 
the size of each FS is very small as clarified by the ARPES 
\cite{FeSe-ARPES6,FeSe-ARPES1,FeSe-ARPES5}
and dHvA \cite{dHvA1,dHvA2} studies.
To reproduce experimental bandstructure of FeSe, we introduce
the additional intra-orbital hopping parameters into $H^0_{\rm FeSe}$,
in order to shift the $d_{xy}$-orbital band [$d_{xz/yz}$-orbital band] 
at ($\Gamma$, M, X) points
by ($0$, $-0.25$, $+0.24$) [($-0.24$, $0$, $+0.12$)] in unit eV;
see Appendix A.
These energy shifts might be induced by the self-energy
\cite{DMFT}.
The constructed FSs in the FeSe model is shown in Fig. \ref{fig:FS} (c).
Since each Fermi pocket is very shallow,
the superconductivity in FeSe could be close to a 
BCS-BEC crossover \cite{BEC}.

In Eq. (\ref{eqn:Ham}),
$H_{\rm M}^U$ is the first-principles screened Coulomb potential
for $d$-orbitals given by the ``constrained-RPA method'' 
\cite{Arita}
given as
\begin{eqnarray}
&&\!\!\!\!\!\!\!\!\!\!
H_{\rm M}^U=
\frac12 \sum_{i,l,m,\s,\s'}
\left\{U_{m,l}n_{i,l\s}n_{i,m,\s'}(1-\delta_{l,m}\delta_{\s,-\s'}) \right.
\nonumber \\
&&\left. +J_{m,l}c_{i,m\s}^\dagger c_{i,l\s}
(c_{i,l\s'}^\dagger c_{i,m\s'}+c_{i,m\s'}^\dagger c_{i,l\s'} \delta_{\s,-\s'}) \right\},
\end{eqnarray}
where $U_{m,l}$ and $J_{m,l}$ are orbital-dependent
Coulomb and Hund's interactions for $d$-electrons, respectively \cite{Arita}.
The averaged intra-orbital Coulomb interaction
$\bar{U}\equiv \frac{1}{5}\sum_{l=1}^5U_{l,l}$ 
and Hund's interactions
$\bar{J}\equiv \frac{1}{10}\sum_{l>m}J_{l,m}$ are 
($\bar{U}$, $\bar{J}$)=($7.21$, $0.681$) for FeSe,
and ($\bar{U}$, $\bar{J}$)=($4.23$, $0.568$) for LaFeAsO in unit eV 
\cite{Arita}.
Thus, the ratio $\bar{J}/\bar{U}=0.0945$ in FeSe is much 
smaller than the ratio $\bar{J}/\bar{U}=0.134$ in LaFeAsO.
Such strong material dependence of ($\bar{U}$, $\bar{J}$) 
is understood as follows: $U_{l,m}$ is strongly screened
by the screening bands (excluding the 8 bands in $H^0_{\rm M}$)
whereas the screening of $J_{l,m}$ is much weak,
and the number of the screening bands is small in FeSe \cite{Arita-private}.
The factor $r(<1)$ in Eq. (\ref{eqn:Ham})
is introduced to adjust the spin fluctuation strength.
The ratio $J_{l,m}/U_{l,m}$ is unchanged by introducing the factor $r$
 \cite{Kuroki-reduction,Misawa}.

The $8\times8$ Green function in the orbital basis is given as
\begin{eqnarray}
{\hat G}(k)=({\hat z}^{-1}i\e_n+\mu-{\hat h}^0_{\rm M}(\k))^{-1},
\label{eqn:Gr}
\end{eqnarray}
where $k=(\k,\e_n=(2n+1)\pi T)$,
${\hat h}^0_{\rm M}(\k)$ is the kinetic term in Eq. (\ref{eqn:Ham-kinetic}),
and ${\hat z}^{-1}\equiv 1-\d{\hat \Sigma}/\d \e|_{\e=0}$
represents the mass-enhancement due to the self-energy at the Fermi level.
Here, we introduce the constant mass-enhancement factor
for $d$-orbital $1/z_l (\ge1)$.
Then, Eq. (\ref{eqn:Gr}) gives the coherent part of the Green function,
which mainly determines the low-energy electronic properties.
In the present study, $r$ and $z_l$ are the fitting parameters.
In FeSe, the orbital order is obtained in 
the real first-principles Hamiltonian ($r\approx1$)
by taking the experimental mass-enhancement factors
$z_l^{-1}\approx4$ into account, as shown later.

The $d$-orbital charge (spin) susceptibilities (per spin)
is given in the following $5^2\times5^2$ matrix form:
\begin{eqnarray}
{\hat \chi}^{c(s)}(\q)= 
{\hat \Phi}^{c(s)}(\q)(1-{\hat \Gamma}^{c(s)}{\hat \Phi}^{c(s)}(\q))^{-1}
\label{eqn:chisc}
\end{eqnarray}
where ${\hat \Phi}^{c(s)}(\q)={\hat \chi}^0(\q)+{\hat X}^{c(s)}(\q)$
is the irreducible susceptibility for the charge (spin) channel.
In the SC-VC theory,
we employ the AL-VC as ${\hat X}^{c,s}(\q)$,
and perform the self-consistent calculation with respect to 
the AL-VC and susceptibilities.
Using the Green function in Eq. (\ref{eqn:Gr}),
the bare susceptibility is
\begin{eqnarray}
\chi^{0}_{l,l';m,m'}(q)= -T\sum_k G_{l,m}(k+q)G_{m',l'}(k),
\end{eqnarray}
where $q=(\q,\w_l=2l\pi T)$.
Also, the AL-VC for the charge susceptibility is given as
\begin{eqnarray}
&&\!\!\!\!\!\!\!\!\!
X_{l,l';m,m'}^{{\rm AL},c}(q)=\frac{T}2\sum_{p}\sum_{a\sim h}
\Lambda_{l,l';a,b;e,f}(q;p)
\nonumber \\
&&\!\!\!\!\!\!\!\!\!
 \ \ \times \{  3{V}_{a,b;c,d}^s(p+q){V}_{e,f;g,h}^s(-p)
+{V}_{a,b;c,d}^c(p+q){V}_{e,f;g,h}^c(-p) \}
\nonumber \\
&&\!\!\!\!\!\!\!\!\!
 \ \ \times \Lambda_{m,m';c,d;g,h}'(q;p) ,
 \label{eqn:AL-c}
\end{eqnarray}
where $p=(\p,\w_m)$, and
${\hat V}^{s,c}(q)\equiv{\hat \Gamma}^{s,c}
+ {\hat\Gamma}^{s,c}{\hat\chi}^{s,c}(q){\hat\Gamma}^{s,c} $.
The three-point vertex ${\hat \Lambda}(q;p)$, 
which gives the coupling between two-magnon and one-orbiton, 
is given as
\begin{eqnarray}
&&\Lambda_{l,l';a,b;e,f}(q;p)
\nonumber \\
&&\ \ \ =-T\sum_{k'}G_{l,a}(k'+q)G_{f,l'}(k')G_{b,e}(k'-p),
\end{eqnarray}
and $\Lambda_{m,m';c,d;g,h}'(q;p)\equiv
\Lambda_{c,h;m,g;d,m'}(q;p)+\Lambda_{g,d;m,c;h,m'}(q;-p-q)$.
We stress that the strong temperature dependence of the 
three-point vertex is significant for realizing the orbital order.
We include all $U^2$-terms without the double counting in order to
obtain quantitatively reliable results. 
Equation (\ref{eqn:AL-c}) means that
the charge AL-VC becomes large in the presence of strong spin fluctuations.
More detailed explanations are presented
in the textbook \cite{Text-SCVC} 

In the present study,
we neglected the spin-channel VCs
since it is expected to be unimportant as discussed in Ref. \cite{Ohno-SCVC}.
In Appendix B, we verify
validity of this simplification in the present model
by performing a time-consuming self-consistent calculation
with respect to both charge- and spin-channel 
Maki-Thompson (MT) and AL-VCs.

Hereafter, we mainly discuss the total spin susceptibility,
$\chi^s(\q) \equiv \sum_{l,m}\chi^s_{l,l;m,m}(\q)$, and the
orbital susceptibilities for $O_{x^2-y^2}=n_{xz}-n_{yz}$,
$\chi^c_{x^2-y^2}(\q) \equiv \chi^c_{2,2;2;2}(\q) + \chi^c_{3,3;3,3}(\q)
-2\chi^c_{2,2;3,3}(\q)$.
The divergence of $\chi^c_{x^2-y^2}(\q)$ at $\q=\bm{0}$
gives rise to the FO order $n_{xz} \ne n_{yz}$.
The charge (spin) Stoner factor $\a_{C(S)}$ is given by the maximum
eigenvalue of ${\hat \Gamma}^{c(s)}{\hat \Phi}^{c(s)}(\q)$
in Eq. (\ref{eqn:chisc}),
and the charge (spin) susceptibility is enlarged in proportion to the
charge (spin) Stoner enhancement factor $S_{C(S)}\equiv (1-\a_{C(S)})^{-1}$.

As explained in Ref. \cite{Ohno-SCVC},
the development of $\chi^c_{x^2-y^2}(\bm{0})$ is mainly induced by the 
diagonal elements of ${\hat \Phi}^{c}$ with respect to $l=2,3$.
If we drop the off-diagonal elements of ${\hat \Phi}^{c}$ approximately,
$\chi^c_{x^2-y^2}(\bm{0})$ is given as
\begin{eqnarray}
\chi^c_{x^2-y^2}(\bm{0}) \approx 2\Phi^c/(1-(1-5J/U)U\Phi^c),
\label{eqn:chiQ}
\end{eqnarray}
where $U \equiv U_{2,2}=U_{3,3}$, $J\equiv J_{2,3}$,
$\Phi^{c}\equiv \chi^0_{l,l;l,l}(\bm{0})+X_{l,l;l,l}^c(\bm{0})$ ($l=2$ or $3$).
Thus, the charge Stoner factor is 
$\a_C =(1-5J/U)U\Phi^c \approx (1-5J/U)(1+UX^c)$,
considering the relation $\chi^0(\q)\approx 1/U$.
Within the RPA ($\Phi^c=\Phi^s=\chi^0$), 
only the spin fluctuations develop 
since the relation $\a_S>\a_C$ is satisfied for $J>0$.
However, the opposite relation $\a_C>\a_S$ is realized if
the relation ${\hat \Phi}^{c}\gg \chi^0$ is satisfied due to the 
charge-channel AL-VC
 \cite{Onari-SCVCS}.

\section{Numerical results for LaFeAsO and FeSe}

\begin{figure}[!htb]
\includegraphics[width=.99\linewidth]{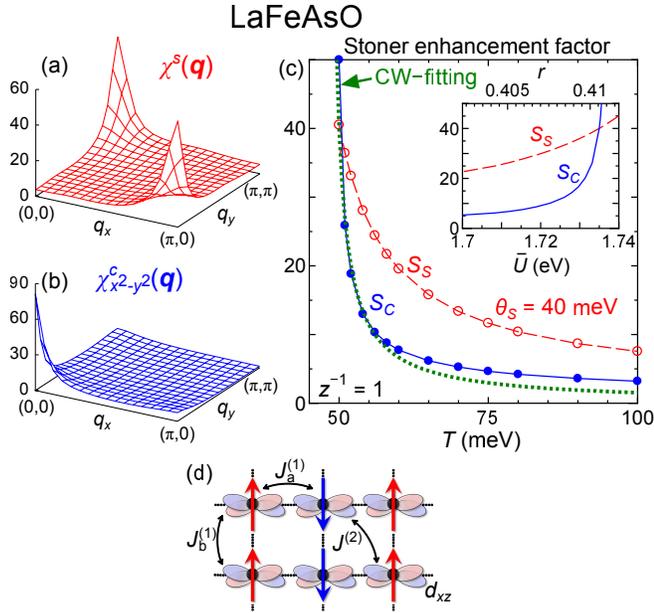}
\caption{
(color online)
(a) $\chi^s(\q)$ and (b) $\chi^c_{x^2-y^2}(\q)$ 
obtained by the SC-VC method for LaFeAsO ($z=1$).
Note that other FO susceptibility ($\chi^c_{z^2}(\q)$)
and antiferro-orbital susceptibilities ($\chi^c_{xz/yz}(\q)$)
also develop secondarily.
(c) Orbital (spin) enhancement factor $S_{C(S)}$
as function of $T$ for LaFeAsO ($z=1$) at $r=0.41$.
The charge and spin Weiss temperatures are 
$\theta_C=48$ meV and $\theta_S=40$ meV, respectively.
Inset: $S_{C(S)}$ as function of $r$ at $T=50$ meV.
Note that $\bar{U}=4.23r$ eV for LaFeAsO.
(d) Orbital-spin interplay in the localized $(d_{xz},d_{yz})$-orbital model,
known as the Kugel-Khomskii coupling.
}
\label{fig:La}
\end{figure}

First, we analyze the LaFeAsO model based on the SC-VC theory.
For $z=1$ for each $l$, the obtained $\chi^s(\q)$ and $\chi^c_{x^2-y^2}(\q)$
are shown in Fig. \ref{fig:La} (a) and (b), respectively,
for $r=0.41$ ($\bar{U}=1.74$ eV) at $T=50$ meV.
Here, the number of $\k$-meshes is $32\times32$, and
the number of Matsubara frequencies is $256$.
Thus, both AFM and FO susceptibilities develop divergently,
and the realized enhancement factors are 
$S_S\approx40$ and $S_C\approx50$.
The $r$-dependences of the enhancement factors at $T=50$ meV
are shown in the inset of Fig. \ref{fig:La} (c):
Both $S_S$ and $S_C$ increase with $r$,
and they are equivalent at $r^*=0.41$.
The lower the temperature is, the smaller $r^*$ is, whereas the value of
$S_S=S_C$ at $r^*$ is approximately independent of $T$.
Similar result is obtained 
in BaFe$_2$As$_2$ model as shown in Appendix C.

In addition, other antiferro-orbital susceptibilities
$\chi^c_{xz}(\q) = 2[\chi^c_{3,4;3,4}(\q) + \chi^c_{3,4;4,3}(\q)]$
and $\chi^c_{yz}(\q) = 2[\chi^c_{2,4;2,4}(\q) + \chi^c_{2,4;4,2}(\q)]$
develop secondary as reported in previous studies 
\cite{Onari-SCVC,Onari-SCVCS,Text-SCVC}.
The obtained results are essentially similar to the results
obtained in the five $d$-orbital Hubbard model for LaFeAsO 
explained in Ref. \cite{Onari-SCVC}.

Figure  \ref{fig:La} (c) shows the temperature dependences
of the Stoner enhancement factors at $r=0.41$.
Both $S_{C}$ and $S_{S}$ 
approximately
follow the Curie-Weiss behaviors with the charge and spin Weiss temperatures
$\theta_C=48$ meV and $\theta_S=40$ meV, respectively.
The obtained relation $\chi^c_{x^2-y^2}(\bm{0})\propto (T-\theta_C)^{-1}$ 
is consistent with the Curie-Weiss behavior 
of the nematic susceptibilities 
in BaFe$_2$As$_2$, derived from $C_{66}$ 
\cite{Yoshizawa,Bohmer},
Raman spectroscopy
\cite{Gallais,Kontani-Raman},
and in-plane resistivity anisotropy 
\cite{Fisher}.
Since $\theta_C\sim\theta_S$, one could interpret that the 
orbital order in LaFeAsO is driven by the spin fluctuations.

The orbital-spin interplay due to the AL-VC is intuitively 
understood in terms of the strong-coupling picture $U\gg W_{\rm band}$
\cite{Onari-SCVCS}:
As shown in Fig. \ref{fig:La} (d),
when the FO order $n_{xz}\gg n_{yz}$ is realized,
the nearest-neighbor exchange interaction has large anisotropy
$J^{(1)}_a\gg J^{(1)}_b$.
Then, the stripe AFM order with $\Q=(\pi,0)$ appears
if $J^{(2)}$ is not too small.
Thus, the FO order/fluctuations and AFM order/fluctuations
emerge cooperatively in the localized model,
and such Kugel-Khomskii-type orbital-spin interplay 
is explained by the AL-VC in the weak-coupling picture.

Next, we analyze the FeSe model,
in which the ratio $\bar{J}/\bar{U}$ is considerably small.
In FeSe, the experimental mass-enhancement factor 
is $\sim10$ for $d_{xy}$-orbital, and $3\sim4$ for other $d$-orbitals
according to the ARPES study \cite{FeSe-ARPES1}.
Therefore, we put $z_l^{-1}=z^{-1}$ for $l\ne4$ and $z_4^{-1}=3z^{-1}$
in the present study.
We find that the peak of $\chi^s(\q)$
moves from $\q=(\pi,\pi)$ to the experimental peak position $\q=(\pi,0)$  
\cite{FeSe-neutron1,FeSe-neutron2,FeSe-neutron3,FeSe-neutron4}
for $z_4^{-1}\ge1.5z^{-1}$,
and the results of the SC-VC method 
are essentially unchanged for $z_4^{-1}\ge1.5z^{-1}$.
Figures \ref{fig:FeSe} (a) and (b) show the 
obtained $\chi^s(\q)$ and $\chi^c_{x^2-y^2}(\q)$ for
$r=0.25$ ($\bar{U}=1.76$ eV) at $T=50$ meV in the case of $z=1$.
We see that only the FO susceptibility develop divergently [$S_C \approx 50$],
whereas the AFM susceptibility remains small [$S_S \approx8$],
consistently with experiments for FeSe.
The $r$-dependences of the Stoner enhancement factors at $T=50$ meV
are shown in the inset of Fig. \ref{fig:FeSe} (c):
With increasing $r$, $S_C$ increases rapidly
whereas $S_S$ remains small.

Figure  \ref{fig:FeSe} (c) shows the temperature dependences
of the enhancement factors at $r=0.25$.
We stress that $S_{C}$ 
{approximately} 
follow the Curie-Weiss 
behavior with the Weiss temperature $\theta_C=48$ meV, which is
consistent with the experimental Curie-Weiss behavior 
with positive $\theta_C$ in FeSe \cite{Ishida-NMR}.
Since the spin Weiss temperature
takes a large negative value ($\theta_S\sim-20$ meV),
which is also consistent with experiments,
one may consider that the orbital order in FeSe 
stems from causes other than spin fluctuations.

\section{Origin of the relation $S_C\gg S_S$ in FeSe}
\label{J-over-U}

In this section, we discuss why the relation $S_C\gg S_S$ 
($\theta_C>0$ and $\theta_S<0$) is realized in FeSe.
First, we focus on the ratio between the Hund's and Coulomb interactions
$\bar{J}/\bar{U}$.
It is intuitively obvious that 
{\it the ratio  $\bar{J}/\bar{U}$ is an important control parameter
for the orbital nematicity}:
For larger $\bar{J}/\bar{U}$, the local configuration 
of the two-electrons in the ($d_{xz},d_{yz}$)-orbitals is
$|d_{xz},\uparrow\rangle  \otimes |d_{yz},\uparrow\rangle$,
where the magnetic moment is $s_z=1$ whereas the 
orbital polarization is $n_{xz}-n_{yz}=0$.
Thus, the smallness of $\bar{J}/\bar{U}$ in FeSe is 
favorable for the emergence of the orbital order without magnetization.

Microscopically, as we discuss in Sec. \ref{Method},
the charge Stoner factor for $\chi^c_{x^2-y^2}(\bm 0)$ is 
$\a_C \approx (1-5J/U)(1+UX^c)$,
where $X^c$ is the charge AL-VC for orbital 2 or 3 at $\q={\bm 0}$.
Since $\bar{J}/\bar{U}=0.0945$ in FeSe,
the orbital order is realized 
by relatively small AL-VC; $X^{c} \sim0.9\chi^0(\bm 0)$.
In LaFeAsO, in contrast, large AL-VC of order
$\sim2\chi^0(\bm 0)$ is required to realize the orbital order.
The obtained AL-VCs in both systems are shown in Fig. \ref{fig:XAL} (c)
in Appendix D.

We discuss why the AL-VC is important in the FeSe model with $\theta_S<0$:
As we explain in Appendix D.
the $T$-dependence of the AL-VC is given as $X^c \sim \Lambda^2 TS_S$,
where $\Lambda$ is the three-point vertex that
represents the interference between two short-living magnons.
We find the relation
$\Lambda^2\propto T^{-a}$ with $a\approx1$ at low temperatures
due to the good nesting between h-FSs and e-FSs \cite{Kontani-Raman}.
Thanks to the strong enhancement of $\Lambda$ at low temperatures,
the orbital order ($\a_C=1$) is realized even if $\theta_S$ is negative.
(Note that $TS_S$ decreases as $T\rightarrow0$ when $\theta_S<0$.)
Thus, serious diagrammatic analysis of the AL-VC is necessary to
understand the rich normal-state phase diagrams in Fe-bases superconductors.

The enhancement of the nematic susceptibility
due to the significant $T$-dependence of $\Lambda^2 (\propto T^{-a})$
had been discussed in Refs. 
\cite{Chubukov-3p,Paul-3p,Kontani-Raman,Khodas-Raman}.
However, the reported exponent $a$ is not universal,
since it depends on the bandstructure and temperature range.
In Appendix E, we show the $T$-dependence of the three-point vertex
for LaFeAsO and FeSe models for wide temperature range.
It is found that $a\approx 1$ for $T=20z \sim 100z [{\rm meV}]$,
where $z<1$ is the band-renormalization factor.
{Due to such large $T$-dependence of $a$, 
$\chi^c_{x^2-y^2}({\bm0})$ obtained by the present study
follows the Curie-Weiss law only approximately.

Finally, we stress the importance of the orbital dependence of 
the spin fluctuation strength.
Since the $d_{xy}$-orbital h-FS (h-FS3) is absent in FeSe,
the relation $\chi^s_{2(3)}(\q) \gg \chi^s_{4}(\q)$
($\chi^s_{l}(\q)\equiv \chi^s_{l,l;l,l}(\q)$) is realized.
This condition is favorable for the development of $\chi^c_{x^2-y^2}(\bm{0})$
since $X^c_{2(3)}$ is enlarged by the spin fluctuations on the 
($d_{xz},d_{yz}$)-orbitals.
More detailed analysis is given in Appendix D.

\begin{figure}[!htb]
\includegraphics[width=.99\linewidth]{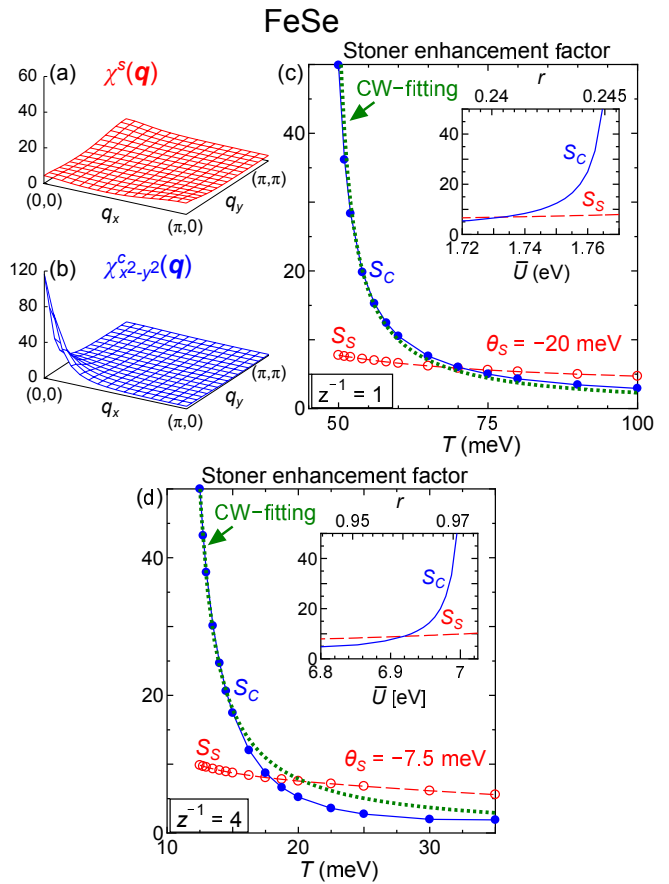}
\caption{
(color online)
(a) $\chi^s(\q)$ and (b) $\chi^c_{x^2-y^2}(\q)$
obtained by the SC-VC method for FeSe ($z=1$),
for $r=0.25$ at $T=50$ meV.
(c) $T$-dependences of the Stoner enhancement factors for FeSe ($z=1$)
at $r=0.25$.
Note that $\bar{U}=7.21r$ for FeSe. 
Inset: Stoner enhancement factors as function of $r$ for FeSe ($z=1$)
at $T=50$ meV.
(d) $T$-dependences of the enhancement factors for FeSe ($z^{-1}=4$)
at $r=0.97$.
The charge and spin Weiss temperatures are $\theta_C=12$ meV
and $\theta_S\sim-7.5$ meV, respectively.
Inset: Stoner enhancement factors as function of $r$ for FeSe ($z^{-1}=4$)
at $T=12.5$ meV.
}
\label{fig:FeSe}
\end{figure}

\section{Effect of the mass-enhancement factor}

Here, we study the effect of the mass-enhancement factor:
We study the FeSe model in the case of $z^{-1}=4$
for ($d_{xz},d_{yz}$)-orbitals.
The obtained $S_{C,S}$ as functions of $r$ are shown in 
the inset of Fig. \ref{fig:FeSe} (d) at $T=12.5$ meV.
Here, $S_S$ remains small even for $r\sim1$
since the bare susceptibility is suppressed by $z$.
In contrast, $S_C$ is enlarged to 50 for $r\approx0.97$,
which is very close to the exact first-principles Hubbard model 
$H_{\rm FeSe}(r=1)$.
The $T$-dependences of $S_{C,S}$
are shown in Fig. \ref{fig:FeSe} (d):
Beautiful Curie-Weiss behavior with $\theta_C=12$ meV
is obtained by putting $r=0.97$.

To understand the similarity between the results in
Fig. \ref{fig:FeSe} (c) for $z=1$ and 
the results in Fig. \ref{fig:FeSe} (d) for $z^{-1}=4$,
we prove that both $\a_{C}$ and $\a_{S}$ are independent of $z$
under the rescaling $T\rightarrow zT$ and $(U,J)\rightarrow (U,J)/z$.
Here, we assume that $z_l^{-1}=z^{-1}$, and
neglect the $T$-dependence of $\mu$ for simplicity.
Under the scaling $T\rightarrow zT$,
the Green function $\hat{G}(\k,n)$
at Matsubara integer $n$ given in Eq. (\ref{eqn:Gr}) is independent of $z$.
For this reason, 
the bare susceptibility $\chi^0(\q)=-T\sum_{\k,n}G(\k+\q,n)G(\k,n)$ 
is proportional to $z$.
By following the same procedure,
the three-point vertex $\Lambda$ is scaled by $z$, and therefore
the AL-VC $X^{c}({\bm 0})\sim TU^4\sum_\q \Lambda({\bm 0};\q)^2\chi^s(\q)^2$
is proportional to $z$
under the scaling $T\rightarrow zT$ and $(U,J)\rightarrow (U,J)/z$.
Thus, both spin and charge irreducible susceptibilities
are proportional to $z$,
and both $\a_S$ and $\a_C$ are unchanged under the rescaling 
$T\rightarrow zT$ and $(U,J)\rightarrow (U,J)/z$.
That is, 
the Weiss temperatures $\theta_{S(C)}$ are scaled by $z$.
The validity of these scaling relations are confirmed 
by the numerical study in Fig. \ref{fig:FeSe}.

It is possible to obtain $z^{-1}$
by calculating the self-energy ${\hat \Sigma}(k)$ 
together with ${\hat \chi}^{s,c}(q)$ self-consistently.
In this case, fine tuning of $r$ will be unnecessary since 
the relation $\a_{S,C}<1$ is assured if ${\hat \Sigma}(k)$ 
and ${\hat \chi}^{s,c}(q)$ are calculated self-consistently 
in two-dimensional systems (Mermin-Wagner theorem) \cite{ROP}.
This is our important future issue.

\section{Discussions}

\subsection{Spin fluctuation strength 
and $\k$-dependent orbital-polarization below $T_{\rm str}$}

Here, we study the electronic states in the FO order
$n_{xz}\ne n_{yz}$ established below the 
structure transition temperature $T_{\rm str}$,
at which the shear modulus $C_{66}$ reaches zero.
According to the linear-response theory, 
$C_{66}\propto 1-g \chi^c_{x^2-y^2}(\bm{0})$,
where $\chi^c_{x^2-y^2}(\bm{0}) \propto (T-\theta_C)^{-1}$ is the 
electronic orbital susceptibility given by the SC-VC theory,
and $g$ is the phonon-mediated Jahn-Teller energy \cite{Kontani-softening}.
Therefore, $C_{66}\propto (T-T_{\rm str})/(T-\theta_C)$, 
and $T_{\rm str}=\theta_C+g$ is slightly higher than $\theta_C$ due to 
the weak electron-phonon coupling ($g\approx10\sim50$ K)
\cite{Yoshizawa,Bohmer,Ishida-NMR}.

Figure \ref{fig:SDW} (a) shows the $T$-dependence of
$S_S$ given by the RPA for LaFeAsO and FeSe for $z=1$.
Here, we introduce the orbital polarization $-\Delta E/2$ ($\Delta E/2$)
for the $d_{xz(yz)}$-level.
We put $S_S=20$ ($5$) for LaFeAsO (FeSe) at $T_{\rm str}=50$ meV,
and assume a mean-field-type $T$-dependence;
$\Delta E= \Delta E_0 {\rm tanh}(1.74\sqrt{T_{\rm str}/T-1})$
with $\Delta E_0=80$ meV.
(For $z^{-1}=4$, the renormalized orbital polarization 
$z \Delta E_0$ is just $20$ meV.)
In both LaFeAsO and FeSe, $S_S$ are enhanced by $\Delta E$,
since $\a_S$ increases {\it linearly} with $\Delta E$ at $\q=(\pi,0)$
as discussed in Ref. \cite{Kontani-softening}.
In LaFeAsO, the magnetic order temperature $T_{\rm mag}$
increases from $\theta_S$ to just below $T_{\rm str}$ 
since $S_S$ is already large at $T_{\rm str}$.
In contrast, in FeSe, the enhancement of $\chi^s(\pi,0)$ is much moderate
\cite{Hirschfeld}.

We also perform the self-consistent analysis of 
the orbital-polarization ($\Delta E_{xz}(\k), \Delta E_{yz}(\k)$)
and anisotropic $\chi^s(\q)$,
which is a natural extension of the SC-VC theory 
into the orbital ordered state
\cite{FeSe-formfactor}.
The obtained $S_S$ and $\k$-dependent orbital polarization 
are shown in Figs. \ref{fig:SDW} (a) (inset) and (b), respectively.
The parameters are $r=0.256$ and $1/z_4=1.6$.
The difference $\Delta n= n_{xz}-n_{yz}$ is 0.2\%.
The hole-pocket around $\Gamma$-point becomes ellipsoidal along the $k_y$-axis
due to the ``sign-reversing orbital polarization'',
in which $\Delta E_{xz}(0,k)-\Delta E_{yz}(k,0)$ 
shows the sign reversal as shown in Fig. \ref{fig:SDW} (c).
Due to this sign reversal,
$S_{S}$ in the inset of Fig. \ref{fig:SDW} (a)
tends to saturate below 40 meV 
\cite{FeSe-ARPES6}.
Also, two Dirac-cone FSs appear around X-point when 
$\Delta E_{yz}(\pi,0)>50$ meV.
These results are essentially
consistent with the recent ARPES studies reported in 
Refs. \cite{FeSe-ARPES1,FeSe-ARPES2,FeSe-ARPES22,FeSe-ARPES3,FeSe-ARPES4,FeSe-ARPES5,FeSe-ARPES6}.
The obtained orbital-polarization ($\Delta E_{xz}(\k), \Delta E_{yz}(\k)$)
belongs to $B_{1g}$ representation, and therefore it is consistent with the 
``$d$-wave orbital order'' discovered in Ref. \cite{FeSe-ARPES4}.
The $d$-wave orbital order is theoretically obtained by the 
mean-field approximation by introducing phenomenological long-range
interaction \cite{FeSe-Hu},
whose microscopic origin might be the AL-VC studied in this paper.

\begin{figure}[!htb]
\includegraphics[width=.9\linewidth]{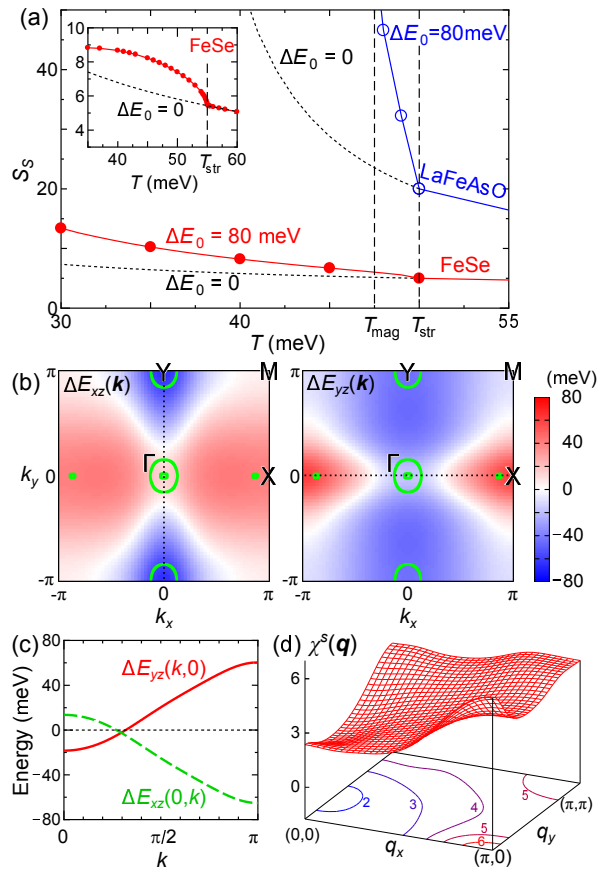}
\caption{
(color online)
(a) $T$-dependences of $S_{S}$ for both LaFeAsO and FeSe models ($z=1$).
The FO order is introduced below $T_{\rm str}=50$ meV.
Inset: $T$-dependences of $S_{S}$ for the FeSe model
obtained by calculating the $\k$-dependent
orbital polarization and $\chi^s(\q)$ self-consistently.
$S_{S}$ tends to saturate below 40 meV
due to the sign-reversing orbital polarization.
(b) Self-consistent solution of the 
orbital polarization ($\Delta E_{xz}(\k), \Delta E_{yz}(\k)$)
in the orbital ordered state in the FeSe model at $T=50$ meV.
The shape of the $C_2$-symmetric FSs in (b) is
consistent with the experimental reports \cite{FeSe-ARPES1,FeSe-ARPES2,FeSe-ARPES22,FeSe-ARPES3,FeSe-ARPES4,FeSe-ARPES5,FeSe-ARPES6}.
We also show 
(c) the $\Delta E_{xz(yz)}(\k)$ along the $k_{y(x)}$-axis, and 
(d) the $C_2$-symmetric $\chi^s(\q)$ in the orbital-ordered state.
}
\label{fig:SDW}
\end{figure}

In the present FeSe model with $z_4^{-1}=3$,
$\chi^s(\q,0)$ has the maximum at $\q=(\pi,0),(0,\pi)$
without orbital order at $T=50$meV,
as shown in Fig. \ref{fig:FeSe} (a).
In the orbital ordered state, $\chi^s(\q,0)$ at $\q=(\pi,0)$ increases
as shown in Fig. \ref{fig:SDW} (d).
These results are consistent with the neutron scattering study for FeSe 
for both $T>T_{\rm str}$ and $T<T_{\rm str}$
\cite{FeSe-neutron1,FeSe-neutron2,FeSe-neutron3,FeSe-neutron4}.
Essentially similar results are obtained for $z_4^{-1}>1.5$ at $T=50$ meV.
We verified that $\chi^s(\q,0)$ has clear maximum peak
at $\q=(\pi,0)$ even for $z_4^{-1}=1.1$ below $T=10$ meV
using $128\times128$ $\k$-mashes.
Experimentally, $z_{2,3}/z_4$ is about three
\cite{FeSe-ARPES1}, 
and the relation $z_{2,3}/z_4 \sim 1.3$ is predicted by the 
dynamical-mean-field-theory in Ref. \cite{DMFT}.

According to Ref. \cite{FeSe-neutron1},
$\chi^s(\q,\w)$ shows the broad maximum at $\q=(\pi,0)$ 
at low-energies ($\w\lesssim10$ meV),
and its strength is almost independent for $T>T_{\rm str}$.
The magnitude of the low-energy spin susceptibility in FeSe 
is one order of magnitude smaller than that in BaFe$_2$As$_2$ \cite{Inosov}, 
whereas its magnitude would be comparable to that in LiFeAs \cite{Qureshi}.
This experimental report in FeSe will be consistent with the 
present theoretical result with the moderate $S_S\sim10$ 
in Figs. \ref{fig:FeSe} (c) and (d).
Note that experimental dispersion relation in $\chi^s(\q,\w)$
for $\w\lesssim 100$ meV is qualitatively understood
based on the present FeSe model by considering the 
band-renormalization factor \cite{FeSe-neutron4}.

\begin{figure}[!htb]
\includegraphics[width=.8\linewidth]{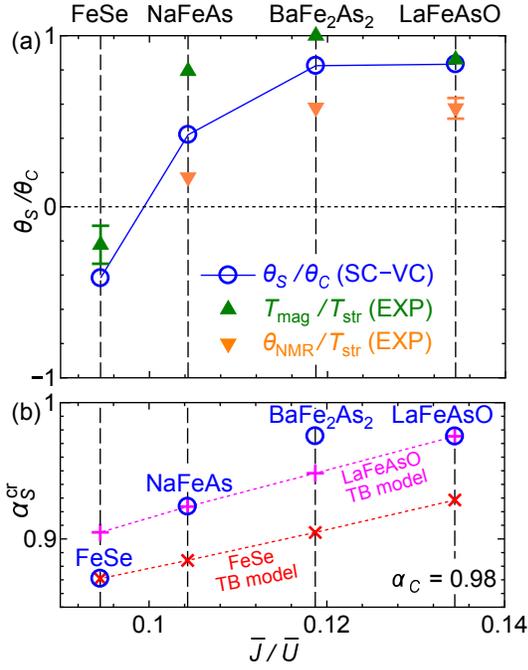}
\caption{
(color online)
(a) Obtained $\theta_S/\theta_C$ for four theoretical models.
Experimental values of $T_{\rm mag}/T_{\rm str}$ and  
$\theta_{\rm NMR}/T_{\rm str}$ are also shown.
$\theta_{\rm NMR}$ is the Weiss temperature of $1/T_1T$
obtained by the Curie-Weiss fitting for $T>T_{\rm str}$ 
in LaFeAsO \cite{NMR-LaFeAsO}, BaFe$_2$As$_2$ \cite{NMR-BaFe2As2}, 
NaFeAs \cite{NMR-NaFeAs}.
In LaFeAsO, we derived $\theta_{\rm NMR}\approx95$ K
from the Curie-Weiss fitting of $1/T_1T$ above $T_{\rm str}$.
In FeSe, we derived $T_{\rm mag}=-10\sim -30$ K from the
Curie-Weiss fitting of $1/T_1T$ below $T_{\rm str}$ \cite{Ishida-NMR}.
Thus, theoretically expected relationships
$\theta_{\rm NMR}/T_{\rm str} \lesssim \theta_S/\theta_C$
and $T_{\rm mag}/T_{\rm str} \gtrsim \theta_S/\theta_C$ are verified.
(b) Obtained $\a_S^{\rm cr}$ for four theoretical models at $\a_C=0.98$.
We also plot $\a_S^{\rm cr}$ for the hybrid models;
$H_{\rm LaFeAsO}^0 + rH_{\rm M}^U$ (LaFeAsO TB model) and
$H_{\rm FeSe}^0 + rH_{\rm M}^U$ (FeSe TB model).
}
\label{fig:ratio}
\end{figure}

\subsection{The ratio $\theta_S/\theta_C$  for FeSe, NaFeAs, BaFe$_2$As$_2$ 
and LaFeAsO as functions of $\bm{{\bar J}/{\bar U}}$ 
\label{sec:ratio}
}
In Fig. \ref{fig:ratio} (a), we summarize the 
ratio $\theta_S/\theta_C$ obtained in
FeSe, NaFeAs, BaFe$_2$As$_2$, and LaFeAsO as function of $\bar{J}/\bar{U}$.
Numerical study for NaFeAs and BaFe$_2$As$_2$
are presented in Appendix C.
In NaFeAs and FeSe, in which ${\bar J}/\bar{U}$ is smaller,
the obtained $\theta_S/\theta_C$ decreases to $0.4$ and $-0.4$, respectively.
In Fig. \ref{fig:ratio} (a),
experimental values of $T_{\rm mag}/T_{\rm str}$ and  
$\theta_{\rm NMR}/T_{\rm str}$ are also shown, where
$\theta_{\rm NMR}$ is the Weiss temperature of $1/T_1T$ above $T_{\rm str}$.
Since $T_{\rm str}= \theta_C+g$ ($g\approx10\sim50$ K)
and $\theta_{\rm NMR}=\theta_S$,
the relation $\theta_{\rm NMR}/T_{\rm str} \lesssim \theta_S/\theta_C$
is expected theoretically.
In addition, the relation $T_{\rm mag}/T_{\rm str} \gtrsim \theta_S/\theta_C$
is expected since $T_{\rm mag}$ is substantially higher than $\theta_S$
in the FO ordered state.
These two theoretically predicted relations
are verified in Fig. \ref{fig:ratio} (a).
Thus, the ratio $\theta_S/\theta_C$ is well scaled by 
the parameter $\bar{J}/\bar{U}$, consistently with the discussion
in Sec. \ref{J-over-U}.

Figure \ref{fig:ratio} (b) shows the critical value of
the spin Stoner factor for $\a_C\approx 1$ in each model, $\a_S^{\rm cr}$.
It is found that $\a_S^{\rm cr}$ increases with $\bar{J}/\bar{U}$ qualitatively.
In addition, we plot $\a_S^{\rm cr}$ for the 
FeSe (LaFeAsO) TB model with different Coulomb interactions:  
$H_{\rm FeSe(LaFeAsO)}^0+ rH_{\rm M}^U$.
In both FeSe and LaFeAsO TB models,
$\a_S^{\rm cr}$ monotonically increases with $\bar{J}/\bar{U}$,
whereas $\a_S^{\rm cr}$ is clearly small for FeSe TB model.
There are two reasons why $\a_S^{\rm cr}$ is smaller 
for the FeSe bandstructure.
One reason is the absence of the $d_{xy}$-orbital h-FS in FeSe:
As we discussed in Sec. \ref{J-over-U},
the $d_{xy}$-orbital spin fluctuations are unnecessary for the 
development of $\chi^c_{x^2-y^2}(\bm{0})$ due to the AL-VC.
Another reason is the smallness of the FSs in FeSe:
We found numerically that $\a_S^{\rm cr}$ decreases
when the size of the FSs is smaller,
since the three-point vertex 
$\Lambda_m \equiv \delta \chi_m^0(\q)/\delta \Delta E_m$, 
which is odd with respect to $G$,
increases in magnitude when the particle-hole asymmetry is large:
In fact, we analyzed the undoped LaFeAsO model with tiny FS pockets
by introducing the positive/negative potentials around the electron/hole FSs, 
and verified that the orbital order is realized by small $\a_S$.
Recently, the advantage of the small FSs for the nematicity
had been stressed by the renormalization group study 
in Ref. \cite{Chubukov-RG}.

\subsection{Summary}
The emergence of the electronic nematic order has attracted 
increasing attention as a fundamental phenomenon in 
strongly correlated metals.
In this paper, we studied the origin of the nematicity 
in Fe-based superconductors, 
by paying the special attention to the  {\it nonmagnetic nematic order} in FeSe.
By applying the orbital+spin fluctuation theory
to the first-principles $d$-$p$ Hubbard models, 
we succeeded in explaining the rich variety of the phase diagrams
in Fe-based superconductors, such as the 
nonmagnetic/magnetic nematic order in FeSe/LaFeAsO. 
The key model parameter to realize rich phase diagram is $J/U$; 
the ratio between the Hund's and Coulomb interactions.
In addition, the ratio $\theta_S/\theta_C$ tends to decrease as the 
size of the FSs shrinks, as discussed in Sec. \ref{sec:ratio}.

In both FeSe and LaFeAsO, strong orbital susceptibility 
$\chi^c_{x^2-y^2}({\bm 0}) \propto (T-\theta_C)^{-1}$
with positive $\theta_C$ is realized by the 
strong orbital-spin interplay due to the strong-coupling effect,
called the Aslamazov-Larkin vertex correction in the field theory.
In the FeSe model, ferro-orbital order is established even when the 
spin Weiss temperature $\theta_S$ is negative as shown in Fig. \ref{fig:FeSe},
since the three-point vertex 
(=the coupling between two-magnon and one-orbiton)
increases at low temperatures as $\Lambda \propto T^{-0.5}$.
In contrast, the spin-nematic susceptibility 
driven by the spin susceptibility should be $T$-independent if $\theta_S<0$, 
as discussed in Ref. \cite{FeSe-neutron4}.
Therefore, we conclude that the nematicity in FeSe 
originates from the orbital order/fluctuations.

The nematic orbital fluctuations might play important roles
in the pairing mechanism in Fe-based superconductors \cite{Kontani-RPA}.
In FeSe, $T_{\rm c}$ increases from 9 K to 40 K under pressure, 
accompanied by the enhancement of spin fluctuations
\cite{FeSe-TS}.
At the same time, the system approaches to the orbital critical point
since $T_{\rm str}$ decreases to zero under pressure.
These facts indicate the important role of the 
spin+orbital fluctuations in FeSe.

\acknowledgments

We are grateful to A. Chubukov, P.J. Hirschfeld, R. Fernandes,
J. Schmalian, Y. Matsuda, T. Shibauchi and T. Shimojima for useful discussions.
This study has been supported by Grants-in-Aid for Scientific 
Research from MEXT of Japan.

\appendix

\section{Eight-orbital models for FeSe and LaFeAsO}

Here, we introduce the eight-orbital $d$-$p$ models for FeSe and LaFeAsO
analyzed in the main text.
We first derived the first principles tight-binding models
using the WIEN2k and WANNIER90 codes.
Crystal structure parameters of FeSe and LaFeAsO 
are given in Refs. \cite{FeSe-t} and \cite{LaFeAsO-to}, respectively.
The obtained bandstructure and FSs in the LaFeAsO model
are shown in Fig. 1 in the main text.
In deriving the FeSe model,
we introduce the $k$-dependent shifts for orbital $l$, $\delta E_l$,
in order to obtain the experimentally observed FSs.
In FeSe, we introduce 
the intra-orbital hopping parameters into $H^0_{\rm FeSe}$
in order to shift the $d_{xy}$-orbital band [$d_{xz/yz}$-orbital band] 
at ($\Gamma$, M, X) points
by ($0$, $-0.25$, $+0.24$) [($-0.24$, $0$, $+0.12$)] in unit eV.
Such level shifts are introduced by the additional 
intra-orbital hopping integrals;
$\delta t^{\rm on-site}=\delta E_\Gamma/4+\delta E_M/4+\delta E_{X}/2$,
$\delta t^{\rm nn}=\delta E_\Gamma/8- \delta E_M/8$, and
$\delta t^{\rm nnn}=\delta E_\Gamma/16+\delta E_M/16-\delta E_{X}/8$.
The bandstructure and FSs in the FeSe model
are shown in Fig. 1 in the main text.

We also explain the orbital-dependent Coulomb interaction.
The bare Coulomb interaction for the spin channel 
in the main text is
\begin{equation}
({\hat \Gamma}^{\mathrm{s}})_{l_{1}l_{2},l_{3}l_{4}} = \begin{cases}
U_{l_1,l_1}, & l_1=l_2=l_3=l_4 \\
U_{l_1,l_2}' , & l_1=l_3 \neq l_2=l_4 \\
J_{l_1,l_3}, & l_1=l_2 \neq l_3=l_4 \\
J_{l_1,l_2}, & l_1=l_4 \neq l_2=l_3 \\
0 , & \mathrm{otherwise}.
\end{cases}
\end{equation}
Also, the bare Coulomb interaction for the charge channel is
\begin{equation}
({\hat \Gamma}^{\mathrm{c}})_{l_{1}l_{2},l_{3}l_{4}} = \begin{cases}
-U_{l_1,l_1}, & l_1=l_2=l_3=l_4 \\
U_{l_1,l_2}'-2J_{l_1,l_2} , & l_1=l_3 \neq l_2=l_4 \\
-2U_{l_1,l_3}' + J_{l_1,l_3} , & l_1=l_2 \neq l_3=l_4 \\
-J_{l_1,l_2} , &l_1=l_4 \neq l_2=l_3 \\
0 . & \mathrm{otherwise}.
\end{cases}
\end{equation}
Here, $U_{l,l}$, $U_{l,l'}'$ and $J_{l,l'}$
are the first principles Coulomb interaction terms
given in Ref. \cite{Arita}.

Finally, we perform the band calculations
for the orthorhombic phase of FeSe and LaFeAsO,
based on the experimental crystal structures.
In both compounds,
the obtained band splitting is too small to explain the 
large orbital polarization ($\sim 60$ meV) observed by ARPES studies.
This result means that the orbital order originate from the 
electron-electron correlation, which is not included in the band calculation.

Figure \ref{fig:ortho} (a) is the 
non-magnetic bandstructure in the orthorhombic LaFeAsO
obtained by the WIEN2k software.
The spin-orbit interaction is not taken into account.
The crystal structure parameters in the orthorhombic phase
is given in Ref. \cite{LaFeAsO-to}.
The orthorhombic structure deformation $(a-b)/(a+b)$ is 0.3\%.
Due to the electron-phonon interaction, 
the four-fold symmetry of the bandstructure is slightly violated:
The splitting between the $d_{xz}$- and $d_{yz}$-bands,
$\Delta E^{\rm band} \equiv E_{yz}-E_{xz}$, is  $16$ meV at X-point,
and $\Delta E^{\rm band}=2$ meV at $\Gamma$-point.

Figure \ref{fig:ortho} (b) is the 
bandstructure in the orthorhombic FeSe.
In the orthorhombic phase, the nearest Fe-Fe length is
$a=2.6716$\r{A} and $b=2.6610$\r{A}, so $(a-b)/(a+b)$ is 0.2\%
\cite{FeSe-t}.
Here, the $\k$-dependent orbital shift 
to fit the ARPES bandstructure introduced above 
is not taken into account.
In FeSe, $\Delta E^{\rm band}=14$ meV at X-point,
and $\Delta E^{\rm band}=3$ meV at $\Gamma$-point.
Thus, the sign reversing orbital splitting observed in 
Ref. \cite{FeSe-ARPES6}
cannot be explained by the band calculation.

The splitting is reduced by the renormalization factor $z$
due to the self-energy.
Since $z\sim1/3$ in FeSe and LaFeAsO,
the renormalized splitting at X-point is 
$z\Delta E^{\rm band}\sim 5$meV,
which is one order of magnitude smaller than the experimental 
orbital splitting.
Therefore, it is confirmed that the origin of the 
electronic nematic state in Fe-based superconductors
is the electron-electron correlation.

\begin{figure}[!htb]
\includegraphics[width=.8\linewidth]{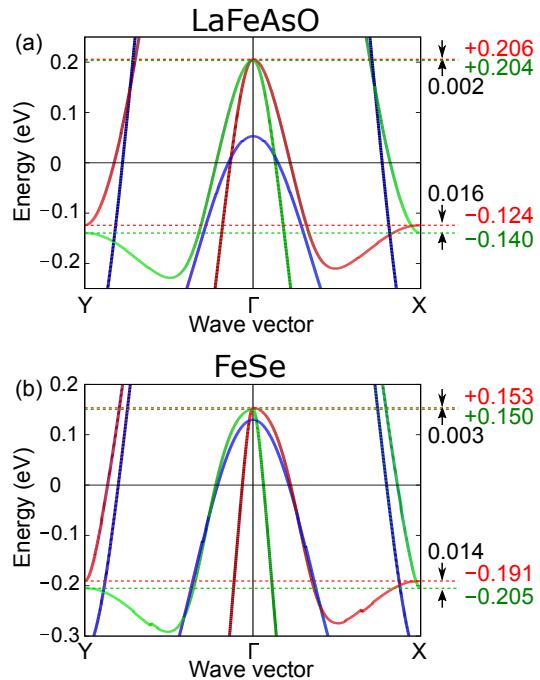}
\caption{
(color online)
Bandstructure of (a) LaFeAsO and (b) FeSe
in the experimental orthorhombic crystal structures
obtained by the WIEN2k software.
}
\label{fig:ortho}
\end{figure}

\section{
Smallness of the VC for the spin susceptibility
}

In the original SC-VC theory,
the spin and charge susceptibilities are calculated self-consistently,
by including the MT-VC and AL-VC for the spin and charge susceptibilities
\cite{Onari-SCVC,Text-SCVC}.
The strong orbital fluctuations are induced by the charge-channel AL-VC 
in Fe-based SCs, Ru-oxides and cuprate SCs
\cite{Onari-SCVC,Onari-SCVCS,Ohno-SCVC}.
In the main text,
we studied the eight-orbital $d$-$p$ Hubbard models 
based on the SC-VC theory,
by taking the charge-channel AL-VC into account self-consistently.
The obtained $\chi^s(\q)$ is equivalent to the RPA 
since the spin-channel VCs are dropped.
It is easy to verify that the charge- and spin-channel MT-VCs 
are negligible in the present model. 
However, the smallness of the spin-channel AL-VC is 
verified only in the two-orbital Hubbard model
in Ref. \cite{Ohno-SCVC}.

Here, we study the FeSe model using the SC-VC method,
by taking the MT-VC and AL-VC for both spin- and charge-channels
in order to confirm the validity of the numerical study in the main text.
The charge (spin) susceptibilities are
\begin{eqnarray}
{\hat \chi}^{c(s)}(\q)= 
{\hat \Phi}^{c(s)}(\q)(1-{\hat \Gamma}^{c(s)}{\hat \Phi}^{c(s)}(\q))^{-1}
\label{eqn:chisc2}
\end{eqnarray}
where ${\hat \Phi}^{c(s)}(\q)={\hat \chi}^{0}(\q)+
{\hat X}^{{\rm MT},c(s)}(\q)+{\hat X}^{{\rm AL},c(s)}(\q)$.
The spin-channel AL-VC is given as
\begin{eqnarray}
&&X_{l,l';m,m'}^{{\rm AL},s}(q)=\frac{T}{2}\sum_p\sum_{a\sim h}\Lambda_{l,l';a,b;e,f}(q;p)
\nonumber \\
&& \ \ \ \times [ \{V^c_{a,b;c,d}(p+q)V^s_{e,f;g,h}(-p)\nonumber\\
&& \ \ \ +V^s_{a,b;c,d}(p+q)V^c_{e,f;g,h}(-p)\}
\Lambda'_{m,m';c,d;g,h}(q;p)\nonumber\\
&& \ \ \ +2V^s_{a,b;c,d}(p+q)V^s_{e,f;g,h}(-p)
\Lambda''_{m,m';c,d;g,h}(q;p)],
\label{eqn:ALs}
\end{eqnarray}
where $\Lambda''_{m,m';c,d;g,h}(q;p)\equiv
\Lambda_{c,h;m,g;d,m'}(q;p)-\Lambda_{g,d;m,c;h,m'}(q;-p-q)$.
Also, the expressions of the charge- and spin-channel MT-VCs 
are given in Ref. \cite{Text-SCVC}.
The double-counting second-order terms with respect to $H^U$
in ${\hat X}^{{\rm MT},s(c)}+{\hat X}^{{\rm AL},s(c)}$
should be subtracted \cite{Text-SCVC} to obtain reliable results.

\begin{figure}[!htb]
\includegraphics[width=.99\linewidth]{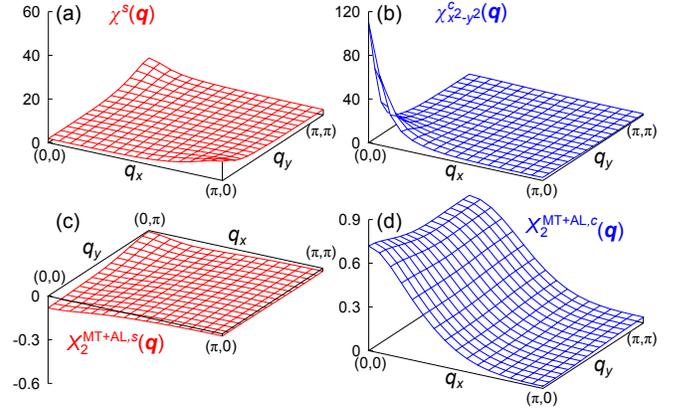}
\caption{
(color online)
(a) $\chi^s(\q)$ and (b) $\chi^c_{x^2-y^2}(\q)$ given by the SC-VC theory,
by calculating both spin- and charge-channel VCs self-consistently.
The obtained results are quantitatively equivalent to Fig. 3 in the main text.
This fact means that the VC for the spin channel is negligible.
(c) ${X}^{{\rm MT+AL},s}_2(\q)$ and (d) ${X}^{{\rm MT+AL},c}_2(\q)$ 
obtained by the present self-consistent calculation.
}
\label{fig:ALS}
\end{figure}

Figures \ref{fig:ALS} (a) and (b) show the 
obtained $\chi^s(\q)$ and $\chi^c_{x^2-y^2}(\q)$, respectively,
for ${\bar U}=1.86$ eV ($r=0.26$) at $T=50$ meV.
The Stoner factors are obtained as $\a_S=0.907$ and $\a_C=0.98$.
The obtained VCs $X^{{\rm MT+AL},s}_2(\q)$ and $X^{{\rm MT+AL},c}_2(\q)$ 
for $d_{xz}$-orbital in the present self-consistent calculation
are shown in Figs. \ref{fig:ALS} (c) and (d), respectively.
Since ${\hat X}^{{\rm MT+AL},s}(\q)$ is very small,
the obtained charge and spin susceptibilities are 
very similar to the results in Fig.3 in the main text.
Therefore, the validity of the numerical results in the main text is confirmed
by performing the very time-consuming self-consistent calculation 
with respect to ${\hat X}^{{\rm MT+AL},s,c}(\q)$ and ${\hat \chi}^{s,c}(\q)$.

\section{Analysis of effective models of BaFe$_2$As$_2$ and NaFeAs
}

In the main text, we introduced the first principles models 
for LaFeAsO and FeSe, and analyzed these models by using the SC-VC method.
Here, we also introduce the effective models for BaFe$_2$As$_2$ and NaFeAs,
and analyze them using the SC-VC method.

In both BaFe$_2$As$_2$ and NaFeAs, the FSs have relatively 
large three-dimensional characters.
In addition, the unfolding of the bandstructure in BaFe$_2$As$_2$
cannot be exactly performed because of its
body-centered tetragonal crystal structure.
Here, we introduce an simple effective BaFe$_2$As$_2$ TB model 
$H^0_{\rm BaFe_2As_2}$ 
by magnifying the size of the $d_{xy}$-orbital hole-FS around $\k=(\pi,\pi)$
in the LaFeAsO unfolded model,
in order to reproduce the ARPES bandstructure in Ba122 compounds.
Here, we shifted the $d_{xy}$-orbital band at M point by $+0.20$ eV.
As for NaFeAs, we just use $H^0_{\rm LaFeAsO}$ 
as an effective NaFeAs TB model, {\it e.g.}, $H^0_{\rm NaFeAs}=H^0_{\rm LaFeAsO}$,
considering that
the FSs in NaFeAs in the $k_z=0$ plane are similar to the FSs in LaFeAsO.
We use $H_{\rm NaFeAs}^U$ in place of $H_{\rm LiFeAs}^U$ 
given in Ref. \cite{Arita}.

\begin{figure}[!htb]
\includegraphics[width=.99\linewidth]{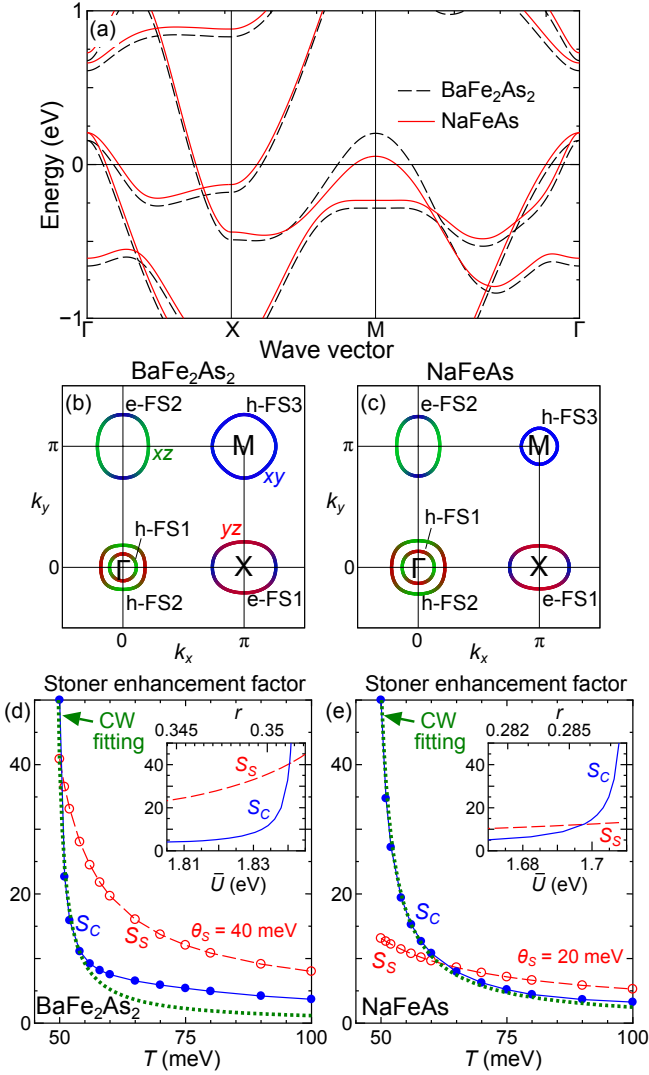}
\caption{
(color online)
(a)  Bandstructures of $H_{\rm BaFe_2As_2}^0$ and $H_{\rm NaFeAs}^0$.
(b) FSs of $H_{\rm BaFe_2As_2}^0$ and (c) FSs of $H_{\rm NaFeAs}^0$.
(d) $T$-dependences of the spin (charge) Stoner enhancement factors
$S_{S(C)}$ obtained in $H_{\rm BaFe_2As_2}$.
(Inset) The $\bar{U}$-dependences of the Stoner enhancement factors.
(e) Spin and charge Stoner enhancement factors in $H_{\rm NaFeAs}$.
}
\label{fig:Ba122}
\end{figure}

The bandstructures and the FSs of the effective TB models 
for BaFe$_2$As$_2$ and NaFeAs are shown in Figs. \ref{fig:Ba122} (a)-(c).
Here, we perform the SC-VC analysis for the models 
$H_{\rm M}=H^0_{\rm M}+ rH^U_{\rm M}$ (M=BaFe$_2$As$_2$, NaFeAs),
where $r(<1)$ is the reduction parameter.
We choose the parameter $r$ to satisfy
the charge Stoner factor is $\a_C=0.98$;
The obtained $T$-dependences of the spin and charge Stoner enhancement factors, 
$S_S\equiv(1-\a_{S})^{-1}$ and $S_C\equiv(1-\a_{C})^{-1}$ respectively,
are shown in Fig. \ref{fig:Ba122} (d) and (e).
As for BaFe$_2$As$_2$,
both spin and orbital fluctuations strongly develop
at $T\sim 50$ meV in the case of $r=0.36$.
This result is consistent with experimental relation 
$T_{\rm mag} \approx T_{\rm str}$ in BaFe$_2$As$_2$.
As for NaFeAs, only orbital fluctuations strongly develop
whereas spin fluctuations remain moderate
at $T\sim 50$ meV in the case of $r=0.287$.
This result is consistent with experimental results in NaFeAs
\cite{NMR-NaFeAs},
in which $T_{\rm mag}(=40{\rm K})$ is more than ten Kelvin smaller than
$T_{\rm str}(=53{\rm K})$.
Thus, normal-state phase diagrams in BaFe$_2$As$_2$ and NaFeAs
are well explained by analyzing their effective Hamiltonians
using the SC-VC method.

\section{Why are strong orbital fluctuations induced by 
tiny spin fluctuations in FeSe?}

In the main text, we studied the
first-principles $d$-$p$ Hubbard models for LaFeAsO and FeSe
by applying the SC-VC theory.
In both models, strong  spin-fluctuation-driven orbital fluctuations
are induced by AL-VC.
In FeSe, we found that
very small spin susceptibility $\chi^s_{\rm max}$ is sufficient to 
realize the orbital order,
consistently with experimental results.

Here, we discuss why strong orbital fluctuations are induced by 
tiny spin fluctuations in FeSe.
In Figs. \ref{fig:XAL} (a) and (b), 
we show the spin and orbital susceptibilities,
$\chi^s_{\rm max}\equiv \chi^s(\Q)$ and $\chi^c_{x^2-y^2}(\0)
\equiv \chi^c_{2,2;2,2}(\q)+\chi^c_{3,3;3,3}(\q)-2\chi^c_{2,2;3,3}(\0)$,
in the FeSe model and LaFeAsO model obtained by the SC-VC theory.
Here, $32\times32$ $\k$-meshes and 256 Matsubara frequencies are used.
In both models, the charge Stoner factor is $\a_C=0.98$
at $T=50$ meV, and the obtained orbital susceptibilities
show similar $T$-dependence.
We set ${\bar U}=1.76$ ($r=0.25$) in FeSe, 
and ${\bar U}=1.74$ ($r=0.41$) in LaFeAsO,
as we did in the main text.
As for the spin susceptibility,
in LaFeAsO, strong spin fluctuations develop at $T=50$ meV ($\a_S=0.98$),
consistently with previous theoretical studies
\cite{Onari-SCVC,Onari-SCVCS}.
In FeSe, in contrast,
$\chi^s_{\rm max}$ is almost constant till $T=50$ meV ($\a_S=0.87$),
consistently with experimental reports in FeSe.

Now, we discuss 
why the spin fluctuation strength required to realize $\a_C\approx1$
is so different from LaFeAsO to FeSe.
One reason is the difference in the ratio $\bar{J}/\bar{U}$:
Figure \ref{fig:XAL} (c) shows the $T$-dependence of 
the AL-VC on $d_{xz}$-orbital,
$X_{2}^{{\rm AL},c}(\bm{0}) \equiv X_{2,2;2,2}^{{\rm AL},c}(\bm{0})$,
obtained in the LaFeAsO and FeSe models.
In both models, $\a_C=0.98$ is satisfied at $T=50$ meV.
At $T=50$ meV, the AL-VC for FeSe is about one-half of that in LaFeAsO.
Thus, small AL-VC is enough
to induce large orbital fluctuations in FeSe,
since the charge Stoner factor is 
$\a_C\approx (1-5{\bar J}/{\bar U})\bar{U}\Phi^c_{2}(\bm{0})$.

In Fig. \ref{fig:XAL} (d), we show that
$X_{2}^{{\rm AL,c}:\ {\rm non-zero}}(\bm{0})
\equiv X_{2}^{{\rm AL},c}(\bm{0})-X_{2}^{{\rm AL},c:\ {\rm zero}}(\bm{0})$
is very small for both FeSe and LaFeAsO.
Here, ``zero'' represents the zero-Matsubara term 
(classical contribution) in Eq. (\ref{eqn:AL-c}) in Sec. \ref{Method}.
Thus, non-zero Matsubara terms in the AL-VC are negligible 
in the present calculation (by chance).
Note that the $U^2$-term in AL-VC gives negative contribution.

\begin{figure}[!htb]
\includegraphics[width=.99\linewidth]{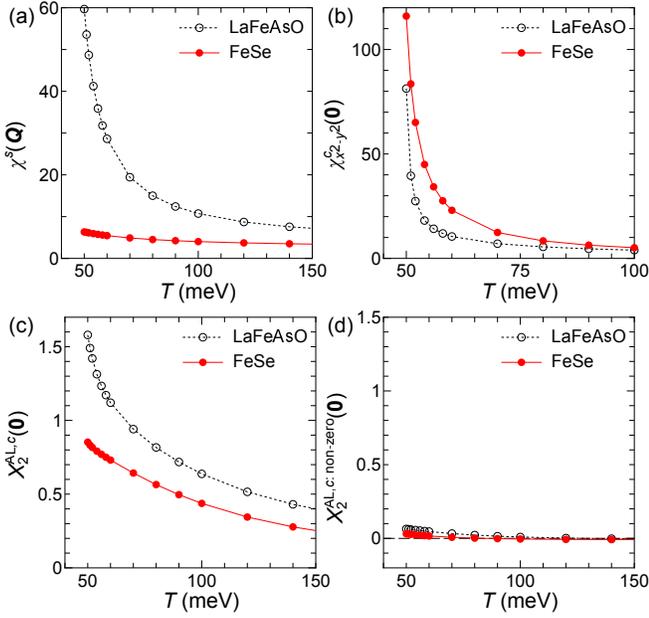}
\caption{
(color online)
(a) Spin susceptibility $\chi^s(\Q)$ and
(b) orbital susceptibilities $\chi^c_{x^2-y^2}(\bm{0})$
for FeSe and LaFeAsO as functions of $T$.
We put $r=0.25$ for FeSe, and $r=0.41$ for LaFeAsO.
(c) $X_{2}^{{\rm AL},c}(\bm{0})$ and 
(d) $X_{2}^{{\rm AL},c:\ {\rm non-zero}}(\bm{0})$ 
for FeSe and LaFeAsO.
}
\label{fig:XAL}
\end{figure}

Another reason for the relation 
$\chi^s_{\rm max}({\rm FeSe})\ll\chi^s_{\rm max}({\rm LaFeAsO})$ 
at $\a_C\approx1$ 
is the difference in the orbital dependence 
of the spin fluctuation strength:
The AL-VC for the $xz$-orbital 
is approximately given as
\begin{eqnarray}
X_{2}^{{\rm AL},c}(\0)&\approx& 3TU^4\sum_{\k}
|\Lambda_{2,2;2,2;2,2}(\0;\k)|^2 \chi_{2}^s(\k)^2
\label{eqn:AL2222}
\end{eqnarray}
where we dropped the inter-orbital terms
of ${\hat \chi}^s$ and ${\hat \Lambda}$,
and leave only the zero-Matsubara term 
in the Matsubara summation in Eq. (\ref{eqn:AL-c}) in Sec. \ref{Method}.
If Eq. (\ref{eqn:AL2222}) is justified, 
only the spin fluctuations on ($d_{xz},d_{yz}$)-orbitals
are important for the FO fluctuations.

\begin{figure}[!htb]
\includegraphics[width=.99\linewidth]{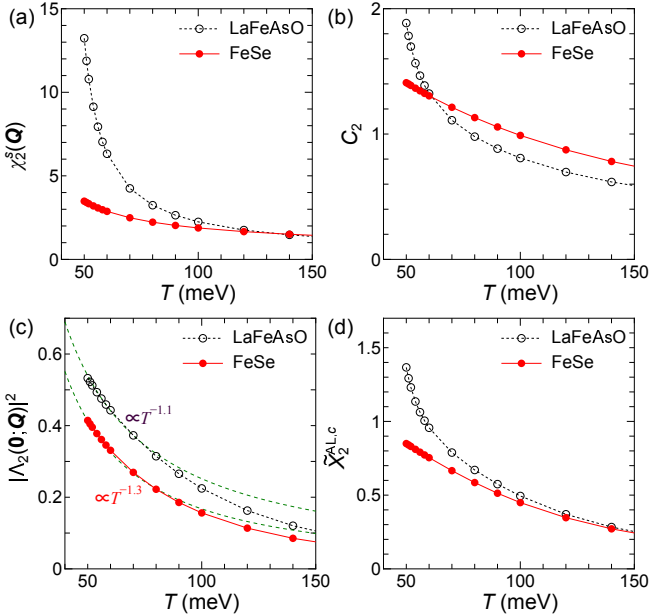}
\caption{
(color online)
(a) $\chi_{2}^s(\Q)$,
(b) $C_2\equiv \sum_\q\chi_{2}^s(\q)^2$, and
(c) $|\Lambda_2|^2$ as functions of $T$ in FeSe and LaFeAsO.
(d) The approximate AL-VC for $d_{xz}$-orbital
${\tilde X}_2^{{\rm AL},c} \equiv 3U^4 |\Lambda_{2}(\0;(0,\pi))|^2 TC_2$.
In both FeSe and LaFeAsO, the
obtained ${\tilde X}_2^{{\rm AL},c}$ qualitatively agrees to
the exact numerical calculations in Fig. \ref{fig:XAL} (c).
}
\label{fig:XAL2}
\end{figure}

Figure \ref{fig:XAL2} (a) shows $\chi_{2}^s(\Q)$ for FeSe and LaFeAsO
for the same model parameters used in Fig. \ref{fig:XAL}.
As derived from Fig. \ref{fig:XAL} (a) and Fig. \ref{fig:XAL2} (a),
the ratio $\chi_{2}^{s}(\Q)/\chi^{s}(\Q)$ is just 0.22 in LaFeAsO,
whereas the ratio increases to 0.53 in FeSe, since the relation
$\chi_{4}^{s}(\Q)\ll\chi_{2}^{s}(\Q)$ ($\chi_{4}^{s}(\Q)\sim\chi_{2}^{s}(\Q)$)
is satisfied in FeSe (LaFeAsO) because of the absence (presence) of h-FS3.
This orbital dependence of the spin fluctuations in FeSe
is favorable for realizing the FO fluctuations.

To understand the model-dependence of the AL-VC in more detail,
we calculate $C_2\equiv \sum_\q\chi_{2}^s(\q)^2$ 
and show the result in Fig. \ref{fig:XAL2} (b):
The ratio $C_2^{\rm LaFeAsO}/C_2^{\rm FeSe}$ is just 1.35
since the width of the peak of $\chi_{2}^s(\q)^2$ around $\q=\Q$ 
is much wider in FeSe.
We also examine the square of the three-point vertex
for $d_{xz}$-orbital $\Lambda_2 \equiv \Lambda_{2,2;2,2;2,2}(\q,\k)$
at $\q=\bm{0}$ and $\k=\Q$ in Fig. \ref{fig:XAL2} (c).
In both models, the relation $|\Lambda_2|^2\propto T^a$ with $a\approx1$
is satisfied for wide temperature range:
Such strong $T$-dependence of the charge-spin coupling $\Lambda_2$
is essential for realizing the orbital fluctuations, so
it should be taken into account in the numerical calculation.
As results, we obtain a crude approximation for the AL-VC,
${\tilde X}_2^{{\rm AL},c} \equiv 3U^4 |\Lambda_{2}(\0;(0,\pi))|^2 TC_2$,
and show the result in Fig. \ref{fig:XAL2} (d).
This crude approximation qualitatively reproduces
the exact numerical results for both FeSe and LaFeAsO 
given in Fig. \ref{fig:XAL} (c).

In summary, in both LaFeAsO and FeSe,
strong orbital fluctuations are induced by AL-VC for the 
$d_{xz(yz)}$-orbital, $X_{2(3)}^{{\rm AL},c}(\bm{0})$.
In FeSe,
very small spin susceptibility $\chi^s_{\rm max}$ is sufficient to 
realize the spin-fluctuation-driven orbital order,
because of both the smallness of $\bar{J}/\bar{U}$
and the largeness of $C_2$.
Strong $T$-dependence of $\Lambda_2$ is essential 
for realizing the orbital fluctuations due to AL-VC.

\section{Strong $T$-dependence of the three-point vertex
}

In this paper, 
we found that the strong orbital fluctuations 
in Fe-based superconductors originate from
the AL-VC for the orbital susceptibility.
The moderate increment of the AL-VC at low temperatures
shown in Fig. \ref{fig:XAL} (c) 
gives rive to the Curie-Weiss behavior of $\chi^c_{x^2-y^2}({\bm 0})$.
For the increment of the AL-VC,
the strong $T$-dependence of the three-point vertex,
shown in Fig. \ref{fig:XAL2} (c),
plays the significant role.
Its strong $T$-dependence in Fe-based superconductors
had been pointed out in Refs. 
\cite{Chubukov-3p,Paul-3p,Kontani-Raman,Khodas-Raman}.

\begin{figure}[!htb]
\includegraphics[width=.7\linewidth]{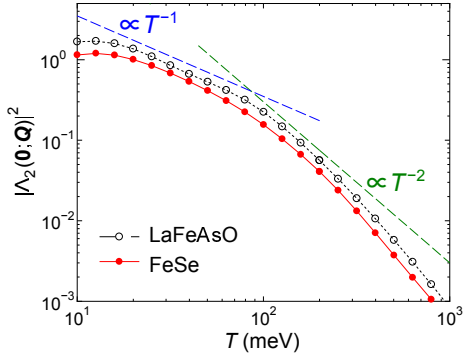}
\caption{
(color online)
$T$-dependences of $|\Lambda_2({\bm 0}; {\bm Q})|^2$
in LaFeAsO and FeSe models.
$512\times512$ $\k$-meshes are used.
}
\label{fig:3point}
\end{figure}

Here, we calculate the three-point vertex for LaFeAsO and FeSe models 
for wide temperature range with high numerical accuracy,
using $512\times512$ $\k$-meshes and $\sim 2048$ Matsubara frequencies.
Figure \ref{fig:3point}
shows the square of the three-point vertex
for $d_{xz}$-orbital $\Lambda_2({\bm 0}; {\bm Q}) 
\equiv \Lambda_{2,2;2,2;2,2}({\bm 0}; {\bm Q})$ for $T\ge10$ meV.
In both LaFeAsO and FeSe models,
the coefficient $a$ of $|\Lambda_2({\bm 0}; {\bm Q})|^2 \propto T^{a}$
depends on the temperature range.
In both models, $a\approx 1$ for $T=20 \sim 100 {\rm meV}$,
so the numerical result in Fig. \ref{fig:XAL2} (c)
is confirmed by this accurate calculation.
When the band renormalization due to $z<1$ is considered,
the relation $a\approx 1$ is realized for $T=20z \sim 100z [{\rm meV}]$.

For $T< 20z$ [meV],
$|\Lambda_2({\bm 0}; {\bm Q})|^2$ saturates 
since the temperature is smaller than the 
energy scale of the nesting.
For $T> 100z$ [meV], the relation $a\approx2$ is realized 
as discussed in Ref. \cite{Chubukov-3p,Paul-3p}.
Note that the chemical potential $\mu$ becomes higher than
the of the hole-band at $\Gamma$ point when $T$ is higher than 
$100z$ ($300z$) [meV] in the FeSe (LaFeAsO) model.

In Ref. \cite{Kontani-Raman}, we reported the relation 
$X_2^{{\rm AL},c}({\bm0})/T\propto T^{-0.5}(1-\a_S)^{-1}$
based on approximate calculation for the five-orbital LaFeAsO TB model.
The factor $T^{-0.5}$ originate from $|\Lambda_2|^2$.
However, we performed more careful numerical analysis, and found that
the approximate relation $|\Lambda_2|^2\sim T^{-1}$ ($a\approx1$) 
is realized for $T\sim50z$ [meV] in the five-orbital model.

\section{Two definitions of the averaged Coulomb and Hund's interactions}

In the present study,
the ratio between the intra-orbital Coulomb interaction
and Hund's interaction, ${\bar J}/{\bar U}$, 
is the essential control parameter.
In this paper, we follow the Hubbard-Kanamori definition:
$\bar{U}\equiv \frac{1}{5}\sum_{l=1}^5U_{l,l}$ and 
$\bar{J}\equiv \frac{1}{10}\sum_{l>m}J_{l,m}$.
This definition is used in Ref. \cite{Arita}.
By using the Slater integrals \cite{ref4-AP}, they are expressed as
$\bar{U} = F^0 + \frac{4}{49} \left(F^2 + F^4 \right)$, 
$\bar{U}' = F^0 - \frac{1}{49} \left( F^2 + F^4 \right)$, and 
$\bar{J} = \frac{5}{98} \left( F^2 + F^4 \right)$
 \cite{ref4-AP}. 
According to the first-principles cRPA method \cite{Arita},
the ratio ${\bar J}/{\bar U}$ is $0.0945$ for FeSe,
and the relation $\bar{U}' = \bar{U} - 2 \bar{J}$ is approximately satisfied.

In the first principles studies,
another definition of the averaged interaction is used frequently:
$\tilde{U} = F^0$ and 
$\tilde{J} = \frac{1}{14} \left( F^2 + F^4 \right)$.
This definition is used in Refs. \cite{ref1-AP,ref2-AP}.
They are equivalent to
$\tilde{U}  = \frac{1}{25} 
\left( \sum_{l=1}^{5} U_{l,l} + \sum_{l \ne m} U'_{l,m} \right)$ and 
$\tilde{J} = \tilde{U}_S - \frac{1}{10} \sum_{l\ne m} 
\left( U'_{l,m} - J_{l,m} \right)$
 \cite{ref5-AP,ref6-AP}.

Therefore, if we assume $\bar{U}' = \bar{U} - 2 \bar{J}$,
which is actually satisfied well in Ref. \cite{Arita}, 
we obtain the relations
$\tilde{U}=\bar{U}-\frac{8}{5} \bar{J}$ and 
$\tilde{J}=\frac{7}{5} \bar{J}$.
Thus, $\tilde{J}/\tilde{U} = 0.224$ obtained by the 
first principles study for FeSe in Ref. \cite{ref2-AP}
corresponds to $\bar{J}/\bar{U}=0.127$.
Also, $\tilde{J}/\tilde{U}=0.294$ for LaFeAsO
obtained in Refs. \cite{ref1-AP,ref2-AP}
corresponds to $\bar{J}/\bar{U}=0.157$.

As shown in Ref. \ref{fig:ratio} (b),
the value of $\a_S^{\rm cr}$ remains small ($\sim0.9$) in the FeSe TB model
with $\hat{H}_{\rm BaFe_2As_2}^U$ (${\bar J}/{\bar U}=0.12$) or 
with $\hat{H}_{\rm LaFeAsO}^U$ (${\bar J}/{\bar U}=0.134$).
In each case, the obtained $T$-dependences of $S_S$ and $S_C$ 
are qualitatively similar to those shown in Fig. \ref{fig:FeSe} (c).
Therefore, the main results of the present study are unchanged
even if $\bar{J}/\bar{U}$ in FeSe is slightly larger than $0.1$.







\begin{thebibliography}{99}


\bibitem{FeSe-TS}
D. C. Johnston, 
{\it The puzzle of high temperature superconductivity in layered iron pnictides and chalcogenides}, 
Adv. Phys. {\bf 59}, 803 (2010);
%
Y. Mizuguchi and Y. Takano, 
{\it Review of Fe Chalcogenides as the Simplest Fe-Based Superconductor.}
J. Phys. Soc. Jpn. {\bf 79}, 102001 (2010).


\bibitem{Ishida-NMR}
A. E. B\"{o}hmer, T. Arai, F. Hardy, T. Hattori, T. Iye, T. Wolf, 
H. v. Lohneysen, K. Ishida, and C. Meingast, 
{\it Origin of the Tetragonal-to-Orthorhombic Phase Transition in FeSe: A Combined Thermodynamic and NMR Study of Nematicity}, 
Phys. Rev. Lett. {\bf 114}, 027001 (2015).


\bibitem{Dresden-NMR}
S.-H. Baek, D. V. Efremov, J. M. Ok, J. S. Kim, Jeroen van den Brink, and B. B\"{u}chner, 
{\it Orbital-driven nematicity in FeSe}, 
Nat. Mater. {\bf 14}, 210 (2015).


\bibitem{FeSe-neutron1}
M. C. Rahn, R. A. Ewings, S. J. Sedlmaier, S. J. Clarke, and A. T. Boothroyd, 
{\it Strong $(\ensuremath{\pi},0)$ spin fluctuations in $\ensuremath{\beta}-\mathrm{FeSe}$ observed by neutron spectroscopy}, 
Phys. Rev. B {\bf 91}, 180501(R) (2015).


\bibitem{FeSe-neutron2}
Q. Wang, Y. Shen, B. Pan, Y. Hao, M. Ma, F. Zhou, P. Steffens, K. Schmalzl, T. R. Forrest, M. Abdel-Hafiez, X. Chen, D. A. Chareev, A. N. Vasiliev, P. Bourges, Y. Sidis, H. Cao, and J. Zhao, 
{\it Strong interplay between stripe spin fluctuations, nematicity and superconductivity in FeSe}, 
Nat. Mater. {\bf 15}, 159 (2016).

\bibitem{FeSe-neutron3}
Q. Wang, Y. Shen, B. Pan, X. Zhang, K. Ikeuchi, K. Iida, A. D. Christianson, 
H. C. Walker, D. T. Adroja, M. Abdel-Hafiez, X. Chen, D. A. Chareev, 
A. N. Vasiliev, and J. Zhao,
{\it Magnetic ground state of FeSe}, 
arXiv:1511.02485.

\bibitem{FeSe-neutron4}
S. Shamoto, K. Matsuoka, R. Kajimoto, M. Ishikado, Y. Yamakawa, T. Watashige, 
S. Kasahara, M. Nakamura, H. Kontani, T. Shibauchi, and Y. Matsuda,
{\it Spin nematic susceptibility studied by inelastic neutron scattering in FeSe}, 
arXiv:1511.04267.

\bibitem{Fernandes}
R. M. Fernandes, L. H. VanBebber, S. Bhattacharya, P. Chandra, V. Keppens, D. Mandrus, M. A. McGuire, B. C. Sales, A. S. Sefat, and J. Schmalian, 
{\it Effects of Nematic Fluctuations on the Elastic Properties of Iron Arsenide Superconductors}, 
Phys. Rev. Lett. {\bf 105}, 157003 (2010).

\bibitem{DHLee}
F. Wang, S. A. Kivelson, and D.-H. Lee, 
{\it Nematicity and quantum paramagnetism in FeSe}, 
Nat. Phys. {\bf 11}, 959 (2015).


\bibitem{Chubukov}
A. V. Chubukov, R. M. Fernandes, and J. Schmalian, 
{\it Origin of nematic order in FeSe}, 
Phys. Rev. B {\bf 91}, 201105(R) (2015).


\bibitem{Mazin}
J. K. Glasbrenner, I. I. Mazin, H. O. Jeschke, P. J. Hirschfeld, R. M. Fernandes, and R. Valent\'{i}, 
{\it Effect of magnetic frustration on nematicity and superconductivity in iron chalcogenides}, 
Nat. Phys. {\bf 11}, 953 (2015). 

\bibitem{QSi}
R. Yu and Q. Si, 
{\it Antiferroquadrupolar and Ising-Nematic Orders of a Frustrated Bilinear-Biquadratic Heisenberg Model and Implications for the Magnetism of FeSe}, 
Phys. Rev. Lett. {\bf 115}, 116401 (2015).


\bibitem{Kruger}
F. Kr\"{u}ger, S. Kumar, J. Zaanen, and J. van den Brink, 
{\it Spin-orbital frustrations and anomalous metallic state in iron-pnictide superconductors}, 
Phys. Rev. B {\bf 79}, 054504 (2009).


\bibitem{PP}
W. Lv, J. Wu, and P. Phillips, 
{\it Orbital ordering induces structural phase transition and the resistivity anomaly in iron pnictides}, 
Phys. Rev. B {\bf 80}, 224506 (2009).

\bibitem{WKu}
C.-C. Lee, W.-G. Yin, and W. Ku, 
{\it Ferro-Orbital Order and Strong Magnetic Anisotropy in the Parent Compounds of Iron-Pnictide Superconductors}, 
Phys. Rev. Lett. {\bf 103}, 267001 (2009).


\bibitem{Onari-SCVC}
S. Onari and H. Kontani, 
{\it Self-consistent Vertex Correction Analysis for Iron-based Superconductors: Mechanism of Coulomb Interaction-Driven Orbital Fluctuations}, 
Phys. Rev. Lett. {\bf 109}, 137001 (2012).


\bibitem{Yoshizawa}
M. Yoshizawa, D. Kimura, T. Chiba, S. Simayi, Y. Nakanishi, K. Kihou, C.-H. Lee, A. Iyo, H. Eisaki, M. Nakajima, and S. Uchida, 
{\it Structural Quantum Criticality and Superconductivity in Iron-Based Superconductor Ba(Fe$_{1-x}$Co$_x$)$_2$As$_2$}, 
J. Phys. Soc. Jpn. {\bf 81}, 024604 (2012).


\bibitem{Bohmer}
A. E. B\"{o}hmer, P. Burger, F. Hardy, T. Wolf, P. Schweiss, R. Fromknecht, M. Reinecker, W. Schranz, and C. Meingast, 
{\it Nematic Susceptibility of Hole-Doped and Electron-Doped ${\mathrm{Ba}\mathrm{F}\mathrm{e}}_{2}{\mathrm{As}}_{2}$ Iron-Based Superconductors from Shear Modulus Measurements}, 
Phys. Rev. Lett. {\bf 112}, 047001 (2014).


\bibitem{Gallais}
Y. Gallais, R. M. Fernandes, I. Paul, L. Chauvi\`{e}re, Y.-X. Yang, M.-A. M\'{e}asson, M. Cazayous, A. Sacuto, D. Colson, and A. Forget, 
{\it Observation of Incipient Charge Nematicity in ${\mathrm{Ba}(\mathrm{Fe}{}_{1\ensuremath{-}X}\mathrm{Co}{}_{X})}_{2}\mathrm{As}{}_{2}$}, 
Phys. Rev. Lett. {\bf 111}, 267001 (2013).


\bibitem{Kontani-Raman}
H. Kontani and Y. Yamakawa, 
{\it Linear Response Theory for Shear Modulus ${C}_{66}$ and Raman Quadrupole Susceptibility: 
Evidence for Nematic Orbital Fluctuations in Fe-based Superconductors}, 
Phys. Rev. Lett. {\bf 113}, 047001 (2014).


\bibitem{Khodas-Raman}
M. Khodas and A. Levchenko, 
{\it Raman scattering as a probe of nematic correlations}, 
Phys. Rev. B {\bf 91}, 235119 (2015).


\bibitem{Schmalian-Raman}
U. Karahasanovic, F. Kretzschmar, T. B\"{o}hm, R. Hackl, I. Paul, Y. Gallais, and J. Schmalian, 
{\it Manifestation of nematic degrees of freedom in the Raman response function of iron pnictides}, 
Phys. Rev. B {\bf 92}, 075134 (2015).


\bibitem{Fisher}
J.-H. Chu, H.-H. Kuo, J. G. Analytis, and I. R. Fisher,
{\it Divergent Nematic Susceptibility in an Iron Arsenide Superconductor}, 
Science {\bf 337}, 710 (2012).


\bibitem{Chu}
H.-H. Kuo, J.-H. Chu, S. A. Kivelson, and I. R. Fisher, 
{\it Ubiquitous signatures of nematic quantum criticality in optimally doped Fe-based superconductors}, 
arXiv:1503.00402.


\bibitem{Pom}
C. J. Halboth, W. Metzner, 
{\it $\mathit{d}$-Wave Superconductivity and Pomeranchuk Instability in the Two-Dimensional Hubbard Model}, 
Phys. Rev. Lett. {\bf 85}, 5162 (2000);
C. Honerkamp, M. Salmhofer, N. Furukawa, T. M. Rice, 
{\it Breakdown of the Landau-Fermi liquid in two dimensions due to umklapp scattering}, 
Phys. Rev. B {\bf 63}, 035109 (2001);
H. Yamase and H. Kohno, 
{\it Instability toward Formation of Quasi-One-Dimensional Fermi Surface in Two-Dimensional t-J Model}, 
J. Phys. Soc. Jpn. {\bf 69} (2000) 2151.

\bibitem{ARPES-Shen}
M. Yi, D. Lu, J.-H. Chu, J. G. Analytis, A. P. Sorini, A. F. Kemper, B. Moritz, S.-K. Mo, R. G. Moore, M. Hashimoto, W.-S. Lee, Z. Hussain, T. P. Devereaux, I. R. Fisher, and Z.-X. Shen, 
{\it Symmetry-breaking orbital anisotropy observed for detwinned Ba(Fe$_{1-x}$Co$_x$)$_2$As$_2$ above the spin density wave transition}, 
Proc. Natl. Acad. Sci. USA {\bf 108}, 6878 (2011).


\bibitem{FeSe-ARPES1}
J. Maletz, V. B. Zabolotnyy, D. V. Evtushinsky, S. Thirupathaiah, A. U. B. Wolter, L. Harnagea, A. N. Yaresko, A. N. Vasiliev, D. A. Chareev, A. E. B\"{o}hmer, F. Hardy, T. Wolf, C. Meingast, E. D. L. Rienks, B. B\"{u}chner, and S. V. Borisenko, 
{\it Unusual band renormalization in the simplest iron-based superconductor ${\text{FeSe}}_{1\ensuremath{-}x}$}, 
Phys. Rev. B {\bf 89}, 220506(R) (2014).


\bibitem{FeSe-ARPES2}
K. Nakayama, Y. Miyata, G. N. Phan, T. Sato, Y. Tanabe, T. Urata, K. Tanigaki, and T. Takahashi, 
{\it Reconstruction of Band Structure Induced by Electronic Nematicity in an FeSe Superconductor}, 
Phys. Rev. Lett. {\bf 113}, 237001 (2014).


\bibitem{FeSe-ARPES22}
M. D. Watson, T. K. Kim, A. A. Haghighirad, N. R. Davies, A. McCollam, A. Narayanan, S. F. Blake, Y. L. Chen, S. Ghannadzadeh, A. J. Schofield, M. Hoesch, C. Meingast, T. Wolf, and A. I. Coldea,
{\it Emergence of the nematic electronic state in FeSe}, 
Phys. Rev. B {\bf 91}, 155106 (2015).


\bibitem{FeSe-ARPES3}
T. Shimojima, Y. Suzuki, T. Sonobe, A. Nakamura, M. Sakano, J. Omachi, K. Yoshioka, M. Kuwata-Gonokami, K. Ono, H. Kumigashira, A. E. B\"{o}hmer, F. Hardy, T. Wolf, C. Meingast, H. v. L\"{o}hneysen, H. Ikeda, and K. Ishizaka, 
{\it Lifting of \textit{xz}/\textit{yz} orbital degeneracy at the structural transition in detwinned FeSe}, 
Phys. Rev. B {\bf 90}, 121111(R) (2014).


\bibitem{FeSe-ARPES4}
P. Zhang, T. Qian, P. Richard, X. P. Wang, H. Miao, B. Q. Lv, B. B. Fu, T. Wolf, C. Meingast, X. X. Wu, Z. Q. Wang, J. P. Hu, and H. Ding, 
{\it Observation of two distinct ${d}_{xz}$/${d}_{yz}$ band splittings in FeSe}, 
Phys. Rev. B {\bf 91}, 214503 (2015).


\bibitem{FeSe-ARPES5}
Y. Zhang, M. Yi, Z.-K. Liu, W. Li, J. J. Lee, R. G. Moore, M. Hashimoto, N. Masamichi, H. Eisaki, S.-K. Mo, Z. Hussain, T. P. Devereaux, Z.-X. Shen, and D. H. Lu, 
{\it Distinctive momentum dependence of the band reconstruction in the nematic state of FeSe thin film}, 
arXiv:1503.01556.


\bibitem{FeSe-ARPES6}
Y. Suzuki, T. Shimojima, T. Sonobe, A. Nakamura, M. Sakano, H. Tsuji, J. Omachi, K. Yoshioka, M. Kuwata-Gonokami, T. Watashige, R. Kobayashi, S. Kasahara, T. Shibauchi, Y. Matsuda, Y. Yamakawa, H. Kontani, and K. Ishizaka, 
{\it Momentum-dependent sign-inversion of orbital polarization in superconducting FeSe}, 
Phys. Rev. B {\bf 92}, 205117 (2015).


\bibitem{Arita}
T. Miyake, K. Nakamura, R. Arita, and M. Imada, 
{\it Comparison of Ab initio Low-Energy Models for LaFePO, LaFeAsO, BaFe$_2$As$_2$, LiFeAs, FeSe, and FeTe: Electron Correlation and Covalency}, 
J. Phys. Soc. Jpn. {\bf 79}, 044705 (2010).


\bibitem{Tsuchiizu-Ru1}
M. Tsuchiizu, Y. Ohno, S. Onari, and H. Kontani, 
{\it Orbital Nematic Instability in the Two-Orbital Hubbard Model: Renormalization-Group + Constrained RPA Analysis}, 
Phys. Rev. Lett. {\bf 111}, 057003 (2013).


\bibitem{Tsuchiizu-Ru2}
M. Tsuchiizu, Y. Yamakawa, S. Onari, Y. Ohno, and H. Kontani,
{\it Spin-triplet superconductivity in ${\mathrm{Sr}}_{2}{\mathrm{RuO}}_{4}$ due to orbital and spin fluctuations: Analyses by two-dimensional renormalization group theory and self-consistent vertex-correction method}, 
Phys. Rev. B {\bf 91}, 155103 (2015)


\bibitem{Onari-SCVCS}
S. Onari, Y. Yamakawa, and H. Kontani, 
{\it High-${T}_{c}$ Superconductivity near the Anion Height Instability in Fe-Based Superconductors: Analysis of ${\mathrm{LaFeAsO}}_{1\ensuremath{-}x}{\mathrm{H}}_{x}$}, 
Phys. Rev. Lett. {\bf 112}, 187001 (2014).


\bibitem{Yamakawa-CDW}
Y. Yamakawa and H. Kontani, 
{\it Spin-Fluctuation-Driven Nematic Charge-Density Wave in Cuprate Superconductors: Impact of Aslamazov-Larkin Vertex Corrections}, 
Phys. Rev. Lett. {\bf 114}, 257001 (2015).

\bibitem{Tsuchiizu-CDW}
M. Tsuchiizu, Y. Yamakawa, and H. Kontani, 
{\it $p$-Orbital Density Wave with $d$ Symmetry in High-$T_c$ Cuprate Superconductors}, 
arXiv:1508.07218.

\bibitem{strong-coupling}
C. Xu, M. M\"{u}ller, and S. Sachdev, 
{\it Ising and spin orders in the iron-based superconductors}, 
Phys. Rev. B {\bf 78}, 020501(R) (2008); 
C. Fang, H. Yao, W.-F. Tsai, J.P. Hu, and S. A. Kivelson, 
{\it Theory of electron nematic order in LaFeAsO}, 
Phys. Rev. B {\bf 77} 224509 (2008); 
E. Abrahams and Q. Si, 
{\it Quantum criticality in the iron pnictides and chalcogenides}, 
J. Phys.: Condens. Matter {\bf 23}, 223201 (2011).

\bibitem{Chubu-Rev}
A. V. Chubukov and P. J. Hirschfeld,
{\it Iron-based superconductors, seven years later}, 
Physics Today, {\bf 68}, 46 (2015);
H. Hosono and K. Kuroki, 
{\it Iron-based superconductors: Current status of materials and pairing mechanism}, 
Physica C {\bf 514}, 399 (2015)


\bibitem{dHvA1}
T. Terashima, N. Kikugawa, A. Kiswandhi, E.-S. Choi, J. S. Brooks, S. Kasahara, T. Watashige, H. Ikeda, T. Shibauchi, Y. Matsuda, T. Wolf, A. E. B\"{o}hmer, F. Hardy, C. Meingast, H. v. L\"{o}hneysen, M.-T. Suzuki, R. Arita, and S. Uji, 
{\it Anomalous Fermi surface in FeSe seen by Shubnikov-de Haas oscillation measurements}, 
Phys. Rev. B {\bf 90}, 144517 (2014).


\bibitem{dHvA2}
A. Audouard, F. Duc, L. Drigo, P. Toulemonde, S. Karlsson, P. Strobel, and A. Sulpice, 
{\it Quantum oscillations and upper critical magnetic field of the iron-based superconductor FeSe}, 
Europhys. Lett. {\bf 109}, 27003 (2015).


\bibitem{DMFT}
Z. P. Yin, K. Haule, and G. Kotliar, 
{\it Kinetic frustration and the nature of the magnetic and paramagnetic states in iron pnictides and iron chalcogenides}, 
Nat. Mater {\bf 10}, 932 (2011);
%
J. Ferber, K. Foyevtsova, R. Valent\'{i}, and H. O. Jeschke, 
{\it LDA $+$ DMFT study of the effects of correlation in LiFeAs}, 
Phys. Rev. B {\bf 85}, 094505 (2012);
%
G. Lee, H. S. Ji, Y. Kim, C. Kim, K. Haule, G. Kotliar, B. Lee, S. Khim, K. H. Kim, K. S. Kim, K.-S. Kim, and J. H. Shim, 
{\it Orbital Selective Fermi Surface Shifts and Mechanism of High ${T}_{c}$ Superconductivity in Correlated $A\mathrm{FeAs}$ ($A=\mathrm{Li}$, Na)}, 
Phys. Rev. Lett. {\bf 109}, 177001 (2012).


\bibitem{BEC}
S. Kasahara, T. Watashige, T. Hanaguri, Y. Kohsaka, T. Yamashita, Y. Shimoyama, Y. Mizukami, R. Endo, H. Ikeda, K. Aoyama, T. Terashima, S. Uji, T. Wolf, H. von L\"{o}hneysen, T. Shibauchi, and Y. Matsuda, 
{\it Field-induced superconducting phase of FeSe in the BCS-BEC cross-over}, 
Proc. Natl. Acad. Sci. USA {\bf 111}, 16309 (2014).

\bibitem{Arita-private}
R. Arita, private communication.

\bibitem{Kuroki-reduction}
K. Suzuki, H. Usui, and K. Kuroki,
{\it Possible Three-Dimensional Nodes in the s$\pm$ Superconducting Gap of BaFe$_2$(As$_{1-x}$P$_x$)$_2$},  
J. Phys. Soc. Jpn. {\bf 80}, 013710 (2011).


\bibitem{Misawa}
T. Misawa and M. Imada, 
{\it Superconductivity and its mechanism in an ab initio model for electron-doped LaFeAsO}, 
Nat. Commun. {\bf 5}, 5738 (2014).

\bibitem{Text-SCVC}
S. Onari and H. Kontani, 
{\it Iron-Based Superconductivity}, 
(ed. P. D. Johnson, G. Xu, and W.-G. Yin, 
Springer-Verlag Berlin and Heidelberg GmbH \& Co. K (2015)).


\bibitem{Ohno-SCVC}
Y. Ohno, M. Tsuchiizu, S. Onari, and H. Kontani, 
{\it Spin-Fluctuation-Driven Orbital Nematic Order in Ru-Oxides: Self-Consistent Vertex Correction Analysis for Two-Orbital Model}, 
J. Phys. Soc. Jpn {\bf 82}, 013707 (2013).


\bibitem{Chubukov-3p}
A. Hinojosa, J. Cai, and A. V. Chubukov, 
{\it Raman resonance in iron-based superconductors: The magnetic scenario}, 
Phys. Rev. B {\bf 93}, 075106 (2016). 

\bibitem{Paul-3p}
I. Paul, 
{\it Nesting-induced large magnetoelasticity in the iron-arsenide systems}, 
Phys. Rev. B {\bf 90}, 115102 (2014).

\bibitem{ROP}
H. Kontani and M. Ohno, 
{\it Effect of a nonmagnetic impurity in a nearly antiferromagnetic Fermi liquid: Magnetic correlations and transport phenomena}, 
Phys. Rev. B {\bf 74}, 014406 (2006). 


\bibitem{Kontani-softening}
H. Kontani, T. Saito, and S. Onari, 
{\it Origin of orthorhombic transition, magnetic transition, and shear-modulus softening in iron pnictide superconductors: Analysis based on the orbital fluctuations theory}, 
Phys. Rev. B {\bf 84}, 024528 (2011). 

\bibitem{Hirschfeld}
S. Mukherjee, A. Kreisel, P. J. Hirschfeld, and B. M. Andersen, 
{\it Model of Electronic Structure and Superconductivity in Orbitally Ordered FeSe}, 
Phys. Rev. Lett. {\bf 115}, 026402 (2015);
A. Kreisel, S. Mukherjee, P. J. Hirschfeld, and B. M. Andersen, 
{\it Spin excitations in a model of FeSe with orbital ordering}, 
Phys. Rev. B {\bf 92}, 224515 (2015). 

\bibitem{FeSe-formfactor}
S. Onari, Y. Yamakawa, and H. Kontani, 
{\it Sign-Reversing Orbital Polarization in the Nematic Phase of FeSe Driven by Aslamazov-Larkin Processes}, 
arXiv:1509.01172.

\bibitem{FeSe-Hu}
K. Jiang, J. Hu, H. Ding, and Z. Wang, 
{\it Interatomic Coulomb interaction and electron nematic bond order in FeSe}, 
Phys. Rev. B {\bf 93}, 115138 (2016).


\bibitem{Inosov}
D. S. Inosov, J. T. Park, P. Bourges, D. L. Sun, Y. Sidis, A. Schneidewind, K. Hradil, D. Haug, C. T. Lin, B. Keimer, and V. Hinkov, 
{\it Normal-state spin dynamics and temperature-dependent spin-resonance energy in optimally doped BaFe$_{1.85}$Co$_{0.15}$As$_{2}$}, 
Nature Physics {\bf 6}, 178 (2010).

\bibitem{Qureshi}
N. Qureshi, P. Steffens, D. Lamago, Y. Sidis, O. Sobolev, R. A. Ewings, L. Harnagea, S. Wurmehl, B. B\"{u}chner, and M. Braden, 
{\it Fine structure of the incommensurate antiferromagnetic fluctuations in single-crystalline LiFeAs studied by inelastic neutron scattering}, 
Phys. Rev. B {\bf 90}, 144503 (2014).

\bibitem{NMR-LaFeAsO}
Y. Nakai, S. Kitagawa, T. Iye, K. Ishida, Y. Kamihara, M. Hirano, and H. Hosono, 
{\it Enhanced anisotropic spin fluctuations below tetragonal-to-orthorhombic transition in LaFeAs(O${}_{1\ensuremath{-}x}$F${}_{x}$) probed by ${}^{75}\phantom{\rule{-0.16em}{0ex}}$As and ${}^{139}$La NMR}, 
Phys. Rev. B {\bf 85}, 134408 (2012).


\bibitem{NMR-BaFe2As2}
F. L. Ning, K. Ahilan, T. Imai, A. S. Sefat, M. A. McGuire, B. C. Sales, D. Mandrus, P. Cheng, B. Shen, and H.-H. Wen, 
{\it Contrasting Spin Dynamics between Underdoped and Overdoped $\mathrm{Ba}({\mathrm{Fe}}_{1-x}{\mathrm{Co}}_{x}{)}_{2}{\mathrm{As}}_{2}$}, 
Phys. Rev. Lett. {\bf 104}, 037001 (2010).


\bibitem{NMR-NaFeAs}
L. Ma, G. F. Chen, D.-X. Yao, J. Zhang, S. Zhang, T.-L. Xia, and W. Yu, 
{\it $^{23}\mathrm{Na}$ and $^{75}\mathrm{As}$ NMR study of antiferromagnetism and spin fluctuations in NaFeAs single crystals}, 
Phys. Rev. B {\bf 83}, 132501 (2011).

\bibitem{Chubukov-RG}
A.V. Chubukov, M. Khodas, and R. M. Fernandes, 
{\it Magnetism, superconductivity, and spontaneous orbital order in iron-based superconductors: who comes first and why?}, 
arXiv:1602.05503.

\bibitem{Kontani-RPA}
H. Kontani and S. Onari, 
{\it Orbital-Fluctuation-Mediated Superconductivity in Iron Pnictides: Analysis of the Five-Orbital Hubbard-Holstein Model}, 
Phys. Rev. Lett. {\bf 104}, 157001 (2010).

\bibitem{FeSe-t}
A. E. B\"{o}hmer, F. Hardy, F. Eilers, D. Ernst, P. Adelmann, P. Schweiss, T. Wolf, and C. Meingast, 
{\it Lack of coupling between superconductivity and orthorhombic distortion in stoichiometric single-crystalline FeSe}, 
Phys. Rev. B {\bf 87}, 180505 (2013).

\bibitem{LaFeAsO-to}
T. Nomura, S. W. Kim, Y. Kamihara, M. Hirano, P. V. Sushko, K. Kato, 
M. Takata, A. L. Shluger, and H. Hosono,
{\it Crystallographic phase transition and high-$T_c$ superconductivity in LaFeAsO:F}, 
Supercond. Sci. Technol. {\bf 21}, 125028 (2008).

\bibitem{ref4-AP}
J. C. Slater, 
{\it The Theory of Complex Spectra}, 
Phys. Rev. {\bf 34}, 1293 (1929).

\bibitem{ref5-AP}
V. I. Anisimov, I. V. Solovyev, M. A. Korotin, M. T. Czy\.{z}yk, and G. A. Sawatzky, 
{\it Density-functional theory and NiO photoemission spectra}, 
Phys. Rev. B {\bf 48}, 16929 (1993).

\bibitem{ref6-AP}
L. Vaugier, H. Jiang, and S. Biermann,
{\it Hubbard $U$ and Hund exchange $J$ in transition metal oxides: Screening versus localization trends from constrained random phase approximation}, 
Phys. Rev. B {\bf 86}, 165105 (2012).

\bibitem{ref2-AP}
M. Aichhorn, S. Biermann, T. Miyake, A. Georges, and M. Imada, 
{\it Theoretical evidence for strong correlations and incoherent metallic state in FeSe}, 
Phys. Rev. B {\bf 82}, 064504 (2010).

\bibitem{ref1-AP}
M. Aichhorn, L. Pourovskii, V. Vildosola, M. Ferrero, O. Parcollet, T. Miyake, A. Georges, and S. Biermann, 
{\it Dynamical mean-field theory within an augmented plane-wave framework: Assessing electronic correlations in the iron pnictide LaFeAsO}, 
Phys. Rev. B {\bf 80}, 085101 (2009).



\end{thebibliography}
\end{document}